%% file: paper.tex
\documentclass[aps,prd,twocolumn,10pt,superscriptaddress,nofootinbib,longbibliography,floatfix]{revtex4-1}

\usepackage[T1]{fontenc}
\usepackage[utf8]{inputenc}
\usepackage{lmodern}
\usepackage{microtype}
\usepackage{amsmath,amssymb,mathtools,bm}
\usepackage{graphicx}
\usepackage{booktabs}
\usepackage{array}
\usepackage{capt-of}
\usepackage{xcolor}
\usepackage{hyperref}
\usepackage{enumitem}

\definecolor{linkblue}{RGB}{24,76,140}
\definecolor{citewine}{RGB}{135,28,50}
\hypersetup{
  colorlinks=true,
  linkcolor=linkblue,
  citecolor=citewine,
  urlcolor=linkblue,
  pdftitle={Transition magnetic-dipole dark matter and the LZ230616 high-recoil candidate},
  pdfauthor={Independent phenomenological study}
}
\setlist{nosep,leftmargin=*}

\newcommand{\mchi}{m_\chi}
\newcommand{\dm}{\delta}
\newcommand{\gM}{g_M}
\newcommand{\mug}{\mu_\gamma}
\newcommand{\sv}{\langle\sigma v\rangle}
\newcommand{\dd}{\mathrm d}
\newcommand{\GeV}{\mathrm{GeV}}
\newcommand{\TeV}{\mathrm{TeV}}
\newcommand{\keV}{\mathrm{keV}}
\newcommand{\cms}{\mathrm{cm^3\,s^{-1}}}

\newcommand{\vmin}{v_{\min}}
\newcommand{\Omegah}{\Omega_\chi h^2}

\newcommand{\Lagr}{\mathcal L}
\newcommand{\eff}{\mathrm{eff}}
\newcommand{\AIFigureNote}{}
\input{sections/fit_macros.tex}
\input{sections/decay_selection_macros.tex}

\input{sections/wide_comparison_macros.tex}

\begin{document}

\title{Transition magnetic-dipole dark matter and the
LZ230616 high-recoil candidate}

\author{Yuxuan He}
\email{he.yx25@cityu.edu.hk}
\affiliation{Department of Physics, City University of Hong Kong, Kowloon, Hong Kong SAR, China}
\date{9 September 2026}

\begin{abstract}
A transition magnetic dipole connects an endothermic nuclear recoil
to a delayed photon. We study this interpretation of LZ230616, the
248-keV candidate in LZ's extended-energy search. Imposing the
thermal abundance in a point hypercharge theory fixes a moment near
$2.4\times10^{-4}\,\GeV^{-1}$ at the preferred masses, leaving a
mass--splitting fit. The excited state travels about 0.6 m, making
its decay relevant to the isolated-recoil selection. A
position-dependent TPC-exit response gives conditional maxima at
$(1.07\,\TeV,346\,\keV)$ without the high-energy sideband and
$(0.44\,\TeV,321\,\keV)$ with its zero-count proxy. Their line rates
exceed the H.E.S.S. Einasto bound by factors of about 11 and 5.5.
We examine halo dependence, other direct and indirect probes, and
charged-messenger matching. Allowing a smaller moment and additional
annihilation preserves the low-mass maxima while admitting heavier
fits over 0.1--100 TeV. Delayed recoil--photon data can measure the
splitting and moment, testing both the detector response and the
assumed connection to thermal annihilation.
\end{abstract}

\maketitle

\input{sections/01_introduction}
\input{sections/02_operator_and_lz}
\input{sections/03_thermal_relic}
\input{sections/04_direct_detection}
\input{sections/05_indirect_cosmology}
\input{sections/06_uv_and_future}
\input{sections/07_conclusions}

\input{sections/note_added}
\input{sections/acknowledgments}

\appendix
\input{sections/appendix_lz_sideband}
\input{sections/appendix_nreft}

\input{sections/appendix_relic}
\input{sections/appendix_lmc}
\input{sections/appendix_solar}
\input{sections/appendix_decay_selection}
\input{sections/appendix_secluded}

\bibliography{references}

\end{document}

%% file: sections/fit_macros.tex
\newcommand{\BZeroMass}{1.081}
\newcommand{\BZeroDelta}{355.1}
\newcommand{\BZeroG}{3.354}
\newcommand{\BZeroMu}{2.349}
\newcommand{\BZeroROI}{0.854}
\newcommand{\BZeroGap}{1.041}
\newcommand{\BZeroSB}{2.263}

\newcommand{\BZeroTau}{0.837}

\newcommand{\BZeroHess}{11.11}

\newcommand{\BZeroCMB}{1.62}
\newcommand{\BZeroCLIC}{2.18}

\newcommand{\BZeroMessenger}{14.2}
\newcommand{\BZeroPZero}{0.104}
\newcommand{\BZeroPZeroMSSI}{0.057}

\newcommand{\BZeroHLDistance}{17.7}
\newcommand{\BOneMass}{0.437}
\newcommand{\BOneDelta}{326.7}
\newcommand{\BOneG}{1.441}
\newcommand{\BOneMu}{2.494}
\newcommand{\BOneROI}{0.755}
\newcommand{\BOneGap}{0.314}
\newcommand{\BOneSB}{0.089}

\newcommand{\BOneTau}{0.954}

\newcommand{\BOneHess}{5.42}

\newcommand{\BOneCMB}{0.84}
\newcommand{\BOneCLIC}{2.39}
\newcommand{\BOneILC}{2.59}
\newcommand{\BOneMessenger}{20.4}
\newcommand{\BOnePZero}{0.915}
\newcommand{\BOnePZeroMSSI}{0.502}

\newcommand{\BOneHLDistance}{5.7}
\newcommand{\LZeroMass}{0.456}
\newcommand{\LZeroDelta}{448.7}
\newcommand{\LZeroG}{1.502}
\newcommand{\LZeroMu}{2.492}
\newcommand{\LZeroROI}{0.829}

\newcommand{\LZeroSB}{6.180}

\newcommand{\LZeroTau}{0.369}

\newcommand{\LOneMass}{0.240}
\newcommand{\LOneDelta}{397.0}
\newcommand{\LOneG}{0.796}
\newcommand{\LOneMu}{2.511}
\newcommand{\LOneROI}{0.775}

\newcommand{\LOneSB}{0.044}

\newcommand{\LOneTau}{0.524}
\newcommand{\LOneLength}{0.331}

\newcommand{\LOneFermiGG}{0.090}

\newcommand{\ThermalNeutralMassPercent}{0.013}
\newcommand{\ThermalInterpolationError}{1.36\times10^{-4}}
\newcommand{\ThermalLateTailError}{2.69\times10^{-7}}
\newcommand{\BZeroXenonWindow}{1.04\times10^{-3}}
\newcommand{\BOneXenonWindow}{1.68\times10^{-3}}

\newcommand{\LOneDeltaCutNineTwentyFive}{387.5}
\newcommand{\LOneMassCutNineTwentyFive}{0.260}
\newcommand{\LOneDeltaCutNineHundred}{376.5}
\newcommand{\LOneMassCutNineHundred}{0.283}
\newcommand{\BZeroSolarCapture}{9.69\times10^{20}}

\newcommand{\BOneSolarCapture}{8.85\times10^{21}}

\newcommand{\ThermalDeltaMin}{280}
\newcommand{\ThermalDeltaMax}{400}
\newcommand{\ThermalDeltaWidthPpm}{4.52}

%% file: sections/decay_selection_macros.tex
\newcommand{\BZeroMeanFlight}{0.600}
\newcommand{\BZeroVolumeEscape}{0.377}
\newcommand{\BZeroEventEscape}{0.397}
\newcommand{\BZeroExitROI}{0.322}
\newcommand{\BZeroPhotonTransparent}{0.325}
\newcommand{\BZeroPhotonEscapeAdded}{0.006}
\newcommand{\BZeroPhotonReturnRemoved}{0.078}

\newcommand{\BZeroPhotonPath}{2.40}

\newcommand{\BOneMeanFlight}{0.602}
\newcommand{\BOneVolumeEscape}{0.379}
\newcommand{\BOneEventEscape}{0.398}
\newcommand{\BOneExitROI}{0.286}
\newcommand{\BOnePhotonTransparent}{0.325}
\newcommand{\BOnePhotonEscapeAdded}{0.005}
\newcommand{\BOnePhotonReturnRemoved}{0.078}

\newcommand{\BOnePhotonPath}{2.15}
\newcommand{\BOneFermiGG}{0.115}
\newcommand{\BOneFermiBoth}{0.150}

\newcommand{\GZeroFermiGG}{1.349}

\newcommand{\GZeroHess}{11.15}

\newcommand{\GOneFermiGG}{0.125}
\newcommand{\GOneFermiBoth}{0.163}
\newcommand{\GOneHess}{5.52}
\newcommand{\GOneAmplitudeLimit}{0.43}

\newcommand{\GZeroMass}{1.072}
\newcommand{\GZeroDelta}{346.3}
\newcommand{\GZeroG}{3.329}
\newcommand{\GZeroROI}{0.834}
\newcommand{\GZeroSideband}{1.520}
\newcommand{\GZeroTau}{0.900}
\newcommand{\GOneMass}{0.442}
\newcommand{\GOneDelta}{321.5}
\newcommand{\GOneG}{1.457}
\newcommand{\GOneROI}{0.745}
\newcommand{\GOneSideband}{0.101}
\newcommand{\GOneTau}{1.001}
\newcommand{\GZeroBufferDelta}{343.4}

\newcommand{\GOneBufferDelta}{319.6}

\newcommand{\GeometryVelocityError}{3.8\times10^{-7}}
\newcommand{\GeometryProbabilityError}{2.7\times10^{-5}}
\newcommand{\GZeroIsolatedTen}{2.94}
\newcommand{\GZeroPairsTen}{2.23}

\newcommand{\GZeroPairRatio}{1.52}
\newcommand{\GOneIsolatedTen}{2.62}
\newcommand{\GOnePairsTen}{2.02}

\newcommand{\GOnePairRatio}{1.54}

%% file: sections/wide_comparison_macros.tex
\newcommand{\WideWOneDropSix}{1.95}
\newcommand{\WideProfileSixDelta}{307.4}
\newcommand{\WideProfileSixROI}{0.472}
\newcommand{\WideProfileSixSideband}{0.530}
\newcommand{\WideProfileSixFlight}{17.7}
\newcommand{\WideProfileSixHess}{0.108}
\newcommand{\WideProfileSixMu}{5.33\times10^{-5}}
\newcommand{\WideProfileSixLine}{4.52\times10^{-28}}

\newcommand{\WideWOneDropFourteen}{1.98}

\newcommand{\WideWOneDropHundred}{2.68}

\newcommand{\WideSplittingError}{4.78\times10^{-5}}

\newcommand{\WideHardExpansionHundred}{9.41}

%% file: sections/01_introduction.tex
\section{Introduction}
\label{sec:introduction}

LZ's extended-energy search reports a nuclear-recoil candidate,
LZ230616, at $248\pm23_{\rm stat}\pm23_{\rm sys}\,\keV$ in
$2.84$ tonne-years \cite{LZ2026HighER}. Endothermic dark matter
naturally favors this energy range: producing a heavier state selects
the fastest incident particles and suppresses low-energy recoils
\cite{TuckerSmithWeiner2001,BramanteEtAl2016}. The candidate offers
a test of this mechanism, although its reported global significance
of $2.6\sigma$ leaves the signal interpretation provisional.

The interpretations available through 9 September 2026 span several
classes. Higgsinos and electroweak multiplets were considered in
Refs.~\cite{FreeseTheodosopoulos2026,FanReece2026,WuZhangZhu2026,
SmirnovEtAl2026}, with high-scale supersymmetric, Peccei--Quinn,
and singlino realizations in
Refs.~\cite{Yin2026,DuWang2026,Visinelli2026,LZEvent260902994,
LZEvent260907811}. Scalar doublets and mixed singlet--doublet
states provide further thermal examples
\cite{Nomura2026,LZEvent260906571,LZEvent260907451,
LZEvent260907800,LZEvent260909138}.
General endothermic and vector-mediated fits were studied in
Refs.~\cite{SuYangYang2026,DiMauro2026,Yamashita2026,
ZhuEtAl2026DarkPhoton}. Related constructions invoke baryon flavor,
gauged baryon number, $B-L$, self-interactions, and extra dimensions
\cite{LZEvent260906171,LZEvent260907225,LZEvent260906909,
LZEvent260906825,LZEvent260907138,LZEvent260909136}.
Their collider implications were also examined
\cite{LZEvent260906750,LZEvent260908712}.

Other proposed origins include exothermic scattering
\cite{DentNewstead2026,deLima2026Exothermic,LZEvent260906153},
elastic spin-dependent scattering \cite{LZEvent260908993},
axion-mediated interactions \cite{Unwin2026,LZEvent260908893},
and boosted particles
\cite{LZEvent260906890,LZEvent260907742,LiangEtAl2026Magnetic}.
Absorption, atmospheric-neutrino upscattering, and bound-neutron
disappearance supply alternatives to a halo recoil
\cite{LouLu2026,JeesunMajumdar2026,LZEvent260909037}.
Annual modulation, nuclear excitation, and heavy-target or
paleo-detector signals provide complementary tests
\cite{McCabe2026,GuEtAl2026Excitation,LZEvent260906640}.
The empty high-energy sideband and solar-neutrino observations
already constrain some interpretations
\cite{RoddEtAl2026Sideband,PospelovRamani2026,
DiMauroShaikh2026Solar,BoseEtAl2026Solar}.
These studies motivate checking the recoil, abundance, and
companion signals within one specified interaction.

For the interaction of a transition magnetic dipole between the Majorana components of a pseudo-Dirac singlet, there is extensive literature
\cite{PospelovTerVeldhuis2000,SigurdsonEtAl2004,
MassoMohantyRao2009,BanksFortinThomas2010,
BargerKeungMarfatia2011,DelNobileEtAl2012Magnetic}.
Magnetic inelastic scattering fixes correlated charge and
nuclear-magnetic responses \cite{ChangWeinerYavin2010},
while the same moment controls coannihilation and radiative decay
\cite{WeinerYavin2012Rayleigh,WeinerYavin2012UV,ArinaEtAl2020}.
The recoil and its delayed photon thus measure one low-energy
interaction. Their spatial and timing information was developed in
Ref.~\cite{LinFinkbeiner2011} and used in a dedicated XENON100 search
\cite{XENON100MiDM2017}. Related work examines smaller splittings,
excited-state cosmology, and Earth-induced photon signals
\cite{ChatterjeeLaha2022,CarrilloGonzalezToro2021,EbyFoxKribs2023}.


We use an electroweak-invariant hypercharge dipole as the reference
theory \cite{ArinaEtAl2020}. It fixes the relative photon and $Z$
moments and relates scattering to coannihilation and neutral-boson
annihilation. We first determine the dipole coupling from the thermal
abundance. The LZ analysis then fits the mass and splitting on this thermal surface,
over a mass domain beginning at $100\,\GeV$.
We report paired benchmarks with and without the empty-sideband proxy
and repeat both fits with a position-dependent excited-state escape
response. This construction keeps the recoil normalization, photon
lifetime, and annihilation rates tied to one coupling. 
We then give an explicit charged-messenger completion, including its
multiplicity and electroweak group factors, and compare published
collider sensitivities with the inferred moment.
An independent annihilation channel provides a second possibility:
it can set the abundance while the magnetic moment controls the recoil
and decay. Appendix~\ref{app:secluded} tests the corresponding
smaller-moment LZ fits and identifies which indirect predictions depend
on the additional interactions.

Section~\ref{sec:model-lz} defines the operator and its high-energy
recoil response. Section~\ref{sec:relic} establishes the thermal surface.
Section~\ref{sec:direct} presents the LZ fits, sideband yields,
recoil--photon selection, and tests with other direct searches.
Indirect and cosmological probes follow in Sec.~\ref{sec:indirect},
and charged-messenger matching and collider comparisons in
Sec.~\ref{sec:uv}. The appendices give the sideband construction,
nonrelativistic matching, relic checks, Large Magellanic Cloud halo
variation, solar capture, detector transport, and the smaller-moment
extension.

%% file: sections/02_operator_and_lz.tex
\section{Transition dipole and high-energy recoils}
\label{sec:model-lz}

A Dirac singlet with small Majorana masses splits into two Majorana
states, $\chi_1$ and $\chi_2$, with masses $\mchi$ and $\mchi+\dm$.
The splitting is protected by the dark-number symmetry restored at
$\dm=0$ \cite{TuckerSmithWeiner2001,CarrilloGonzalezToro2021}.
We take its leading Standard Model interaction to be the hypercharge
Pauli term \cite{ArinaEtAl2020},
\begin{equation}
 \Lagr_B=\frac{C_M}{2\Lambda}\bar\Psi\sigma^{\mu\nu}\Psi B_{\mu\nu}.
 \label{eq:hypercharge-dipole}
\end{equation}
In the mass basis its photon coupling is
\begin{equation}
 \Lagr_\gamma=-\frac{i\mug}{2}\bar\chi_2\sigma^{\mu\nu}\chi_1F_{\mu\nu},
 \qquad \mug=\frac{C_Mc_W}{\Lambda}=\frac{e\gM}{4\mchi}.
 \label{eq:transition-dipole}
\end{equation}
We use the transition vertex $\mug\sigma^{\mu\nu}q_\nu$,
electroweak invariance fixes $\mu_Z=-\mug\tan\theta_W$.
The mostly singlet state is assumed to have subleading Higgs, anapole,
and other interactions at recoil momenta. Section~\ref{sec:uv}
examines the ultraviolet (UV) assumptions behind this hierarchy.

For endothermic scattering on a nucleus of mass $m_T$, the minimum
incident speed and its kinematic minimum are
\cite{TuckerSmithWeiner2001,BramanteEtAl2016}
\begin{align}
 \vmin(E_R)&=\frac{m_TE_R/\mu_{\chi T}+\dm}{\sqrt{2m_TE_R}},
 \label{eq:vmin-main}\\
 E_R^*&=\frac{\mu_{\chi T}}{m_T}\dm,
 \qquad v_* =\sqrt{\frac{2\dm}{\mu_{\chi T}}}.
 \label{eq:kinematic-apex}
\end{align}
Here $\mu_{\chi T}$ is the reduced mass. A splitting near
$0.3$--$0.4$ MeV therefore selects xenon recoils of a few hundred keV.
At the sideband-preferred reference point B1 below,
$E_R^*\simeq255\,\keV$ and $\vmin(248\,\keV)\simeq785$ km/s:
the incident population lies near the halo endpoint.

Photon exchange couples to nuclear charge and magnetism. In the
nonrelativistic basis of
Refs.~\cite{FitzpatrickEtAl2013,AnandFitzpatrickHaxton2013,BarelloChangNewby2014},
the nonzero coefficients are
\begin{align}
 c_1^p&=\frac{e\mug}{2\mchi},&
 c_5^p&=-\frac{2e\mug m_N}{q^2},\nonumber\\
 c_4^{p,n}&=\frac{g_{p,n}e\mug}{m_N},&
 c_6^{p,n}&=-\frac{g_{p,n}e\mug m_N}{q^2},
 \label{eq:nr-matching}
\end{align}
where the nucleon magnetic factors are $(g_p,g_n)=(5.59,-3.83)$
\cite{TiesingaEtAl2021}. We route $\bm q$ toward $\chi_2$ and use
the reference mass $m_N=0.9315\,\GeV$ of the nuclear responses.
Appendix~\ref{app:nreft} defines the operators, derives their relative
signs, and checks the amplitude with exact spinors. All four terms
are combined coherently in the rate,
\begin{equation}
 \frac{\dd R}{\dd E_R}=\frac{\rho_\chi}{\mchi}
 \sum_T\frac{\xi_T}{m_T}\int_{v>\vmin}\dd^3v\,
 f_{\rm lab}(\bm v,t)v\frac{\dd\sigma_T}{\dd E_R}.
 \label{eq:rate-master-main}
\end{equation}
We use natural-xenon mass fractions and one-body nuclear responses
\cite{MeijaEtAl2016,KavanaghWIMpy}. Our Standard Halo Model (SHM)
has $\rho_\chi=0.30\,\GeV\,\mathrm{cm}^{-3}$, $v_0=238$ km/s,
$v_{\rm esc}=544$ km/s, and a 24-phase annual average at SURF
\cite{BaxterEtAl2021,McCabe2014Velocity}.
The selected count folds Eq.~\eqref{eq:rate-master-main} with the
LZ efficiency and $2.84$ tonne-year exposure. 
Appendix~\ref{app:lz-sideband} specifies the response used there.

LZ's elastic magnetic template, $\mathcal L_{10}^{s}$, provides a
related comparison \cite{LZ2026HighER}. The superscript denotes
isoscalar nucleon couplings, and the index labels the relativistic
contact tensor--tensor interaction in
Ref.~\cite[Table~I]{AnandFitzpatrickHaxton2013}. Its reduction is
proportional to $q^2O_4-m_N^2O_6$ and gives a broad, double-peaked
elastic spectrum. The photon pole, charge response, and positive
splitting produce a distinct recoil spectrum in the electromagnetic
transition considered here. The boosted contact study of
Ref.~\cite{LiangEtAl2026Magnetic} uses the former tensor interaction.

\subsection{Radiative decay}
\label{sec:decay-main}

The same moment fixes a second signal
\cite{ChangWeinerYavin2010,LinFinkbeiner2011}:
\begin{equation}
 \Gamma_2=\frac{\mug^2}{8\pi}\frac{(m_2^2-m_1^2)^3}{m_2^3}
 \simeq\frac{\mug^2\dm^3}{\pi}.
 \label{eq:radiative-width}
\end{equation}
The photon has energy $E_\gamma\simeq\dm$, its parent travels a mean
distance $\lambda_2=v_2\tau_2$. The thermal moments found next give
microsecond lifetimes and flight lengths comparable to the LZ chamber.
Section~\ref{sec:decay-selection} includes this decay in the
isolated-recoil fit.

%% file: sections/03_thermal_relic.tex
\section{The thermal relation}
\label{sec:relic}

We carry out the abundance assuming a symmetric relic in a standard radiation-dominated universe, chemical equilibrium with the Standard Model, with the annihilation governed by the point hypercharge operator Eq. \ref{eq:transition-dipole}.
Conversions maintain a common chemical potential for the two states
\cite{CarrilloGonzalezToro2021,ChatterjeeLaha2022}. Since
$\dm/T_f\lesssim2\times10^{-4}$, their freeze-out populations are
nearly equal. For the total density $n=n_1+n_2$, the ordered
coannihilation sum gives \cite{GriestSeckel1991}
\begin{equation}
 \sv_{\eff}=\tfrac12\sv_{12}+\tfrac12\sv_N^{(2)}.
 \label{eq:sigmaeff-degenerate}
\end{equation}
The first term contains $\chi_1\chi_2\to f\bar f,W^+W^-,Zh$,
the second contains the two-insertion neutral-boson channels of either
identical-state pair. The factors of one half follow from the
total-density convention and recover the Dirac limit
\cite{WeinerYavin2012Rayleigh,SteigmanEtAl2012,ArinaEtAl2020}.

In the high-mass, zero-velocity limit,
\begin{equation}
 (\sigma v)_{\eff}=
 \frac{41\alpha\mug^2}{16c_W^4}
 +\frac{\mug^4\mchi^2}{8\pi c_W^4}.
 \label{eq:moment-thermal-scaling}
\end{equation}
The first term fixes an approximately
mass-independent physical moment, while the second becomes important at
multi-TeV masses. Photon--$Z$ interference is essential to the
longitudinal $W$ amplitude \cite{ArinaEtAl2020}.

The numerical calculation retains finite-mass one-insertion kernels
and the finite-mass neutral $s$ wave with its leading velocity correction.
We thermally average the former using the invariant integral of
Ref.~\cite{GondoloGelmini1991} and integrate the yield with the
temperature-dependent Standard Model equation of state
\cite{SaikawaShirai2018}, including its entropy derivative
\cite{DreesEtAl2015}. Appendix~\ref{app:relic} gives the kernels,
normalization, and accuracy checks. These details matter near the
$W$, $Zh$, and top thresholds below 200 GeV.

\begin{figure*}[t]
 \includegraphics[width=\textwidth]{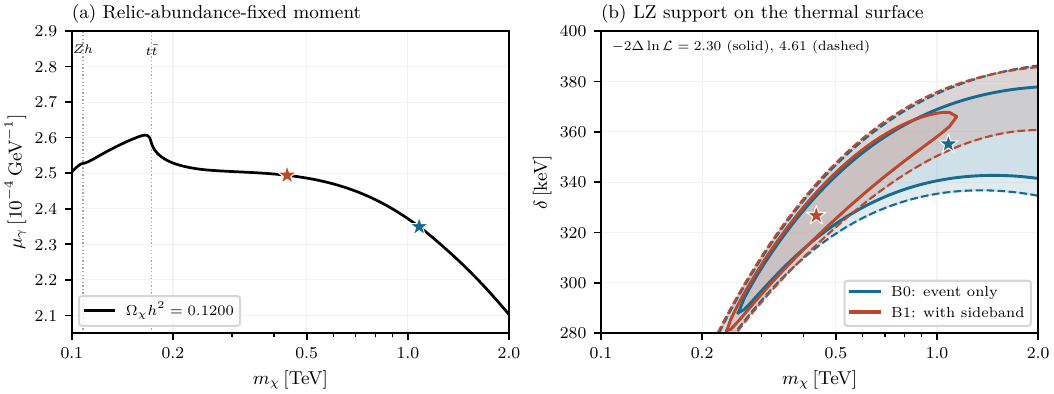}
 \caption{Thermal moment and recoil-only LZ fit. Left: the abundance-fixed moment. Right:
 mass--splitting support with the moment eliminated and energy scale
 profiled. Blue omits the sideband and orange includes its empty
 350--590-keV proxy. Stars mark B0/B1. Solid and dashed contours give
 $-2\ln(\mathcal L/\mathcal L_{\max})=2.30,4.61$, these are
 conditional likelihood levels with uncalibrated coverage.
 The displayed domain is 100 GeV--2 TeV, the reference fit extends
 to 5 TeV. Escape-selected fits follow in Sec.~\ref{sec:decay-selection}. \AIFigureNote}
 \label{fig:thermal-lz-target}
\end{figure*}

We impose the measured abundance \cite{Planck2018} before fitting LZ:
\begin{equation}
 \Omegah(\mchi,\dm,\gM^{\rm th})=0.1200,
 \qquad \gM=\gM^{\rm th}(\mchi,\dm).
 \label{eq:thermal-elimination}
\end{equation}
For masses of a few hundred GeV to about a TeV this gives
$\mug\simeq(2.4$--$2.5)\times10^{-4}\,\GeV^{-1}$, or
$\gM\simeq3.3(\mchi/\TeV)$. Figure~\ref{fig:thermal-lz-target}
shows the thermal relic abundance-fixed moment in the left panel. For mass splitting in $\ThermalDeltaMin$--$\ThermalDeltaMax$ keV interval
of the right panel, its fractional width is only
$\ThermalDeltaWidthPpm\times10^{-6}$: the small splitting barely
changes freeze-out, while it strongly changes the recoil threshold.
This difference produces the narrow, mass-dependent recoil band.

All fits take $\chi_1$ to supply the full dark-matter density.
An underabundant dipole relic needs the corresponding local-density
rescaling. In the one-insertion regime it gives the approximate
coupling cancellation discussed in Ref.~\cite{WeinerYavin2012Rayleigh}.
The alternative with additional annihilation and a smaller moment is
examined in Appendix~\ref{app:secluded}. Both extensions require their
own abundance assumptions. Likewise, resolving the charged states
that generate the dipole changes the hard rates and hence the
thermal relation, as discussed in Sec.~\ref{sec:uv}.

%% file: sections/04_direct_detection.tex
\section{LZ spectra, sideband, and photon selection}
\label{sec:direct}

\subsection{Conditional likelihood and recoil benchmarks}

LZ's extended-energy search uses 220 live days and a $2.84$ tonne-year
exposure. Its candidate has $S1c=540.1$ phd, $S2c=9268$ phd, and
$E_{\rm obs}=248\pm23_{\rm stat}\pm23_{\rm sys}\,\keV$
\cite{LZ2026HighER}. The efficiency is close to 96\% between 14 and
250 keV, with 50\% endpoints at 5.4 and 269.9 keV. The high-energy sideband contains zero events, the simulated
multiple-scintillation, single-ionization background component is
0.6 events \cite{LZ2026HighER,RoddEtAl2026Sideband}.
 The remaining displayed
events contribute a parameter-independent background factor in this
approximation. The candidate's statistical width and common scale pull
$\zeta$ are treated separately:
\begin{equation}
 \bar E=(1+\tfrac{23}{248}\zeta)E_R,
 \qquad \sigma_{\rm stat}(\bar E)=23\sqrt{\frac{\bar E}{248\,\keV}}\,\keV.
 \label{eq:profile-energy-response}
\end{equation}
The reference selected intensity is
\begin{equation}
 \lambda(E)=\mathcal E\int\dd E_R\,
 \frac{\dd R_{\rm th}}{\dd E_R}\epsilon_{\rm LZ}(\bar E)
 G_{\rm ROI}(E|\bar E),
 \label{eq:selected-intensity}
\end{equation}
where $G_{\rm ROI}$ is normalized within the approximate search
window (region of interest, ROI), so that $\int\lambda\,\dd E=N_{\rm ROI}$.
The public efficiency and reconstructed-energy endpoints define a
response approximation. Appendix~\ref{app:lz-sideband} gives its
normalization and variations.

For the sideband we use a unit-response recoil-energy proxy over
350--590 keV, motivated by Ref.~\cite{RoddEtAl2026Sideband}.
The likelihood is
\begin{align}
 \ln\mathcal L_A={}&\ln\lambda(248\,\keV)-N_{\rm ROI}
 -A_{\rm sideband}N_{\rm sideband}\nonumber\\
 &-\tfrac12\zeta^2+\mathrm{constant}.
 \label{eq:thermal-lz-likelihood}
\end{align}
A fixed background in the empty sideband changes only the constant. 

The two best-fitted benchmark points in the abundance-fixed moment curve, 
B0 uses $A_{\rm sideband}=0$ and B1 uses $A_{\rm sideband}=1$.
The unreported gap between the search window and sideband enters only
as a prediction. Each trial point obeys Eq.~\eqref{eq:thermal-elimination}.
We maximize over $0.1\leq\mchi/\TeV\leq5$,
$250\leq\dm/\keV\leq410$, and $|\zeta|\leq3$.
The contours show the profiled likelihood drop
$q=-2\ln(\mathcal L/\mathcal L_{\max})$.
With one assigned event and an approximate response, their confidence
coverage requires calibration \cite{CowanEtAl2011}. LZ's reported
$3.4\sigma$ local and $2.6\sigma$ global significances belong to its
own templates and background likelihood \cite{LZ2026HighER}.

\begin{table*}[t]
\caption{Thermal benchmarks with a common abundance, $\Omegah=0.1200$.
Suffix 0 uses the candidate alone, and suffix 1 adds the empty sideband.
B0/B1 use the recoil response, and G0/G1 additionally require
excited-state exit from the TPC. Counts use the same physical exposure
and the stated sideband proxy. Digits identify the numerical maxima, while
halo and detector uncertainties exceed the numerical precision.}
\label{tab:direct-benchmarks}
\label{tab:geometry-benchmarks}
\begin{ruledtabular}
\begin{tabular}{lccccccc}
Point & $\mchi$ [TeV] & $\dm$ [keV] & $g_M$ &
$\mug$ [$10^{-4}\,\GeV^{-1}$] & $N_{\rm ROI}$ &
$N_{\rm sideband}$ & $\tau_2$ [$\mu$s]\\
\midrule
B0 & \BZeroMass & \BZeroDelta & \BZeroG & \BZeroMu & \BZeroROI & \BZeroSB & \BZeroTau\\
B1 & \BOneMass & \BOneDelta & \BOneG & \BOneMu & \BOneROI & \BOneSB & \BOneTau\\
G0 & \GZeroMass & \GZeroDelta & \GZeroG & 2.352 & \GZeroROI & \GZeroSideband & \GZeroTau\\
G1 & \GOneMass & \GOneDelta & \GOneG & 2.493 & \GOneROI & \GOneSideband & \GOneTau\\
\end{tabular}
\end{ruledtabular}
\end{table*}

The B0 maximum favors a hard spectrum with $\BZeroSB$ sideband
events. B1 moves to a smaller mass and splitting, reducing that
population to $\BOneSB$ while retaining $\BOneROI$ selected events
(Table~\ref{tab:direct-benchmarks}). The zero-count probabilities
are $\BZeroPZero$ and $\BOnePZero$ for zero background. Including the
fixed 0.6-event component gives $\BZeroPZeroMSSI$ and
$\BOnePZeroMSSI$. These are conditional predictive probabilities.

Figure~\ref{fig:decay-spectra} shows both spectra and their integrated
yields. Their shapes reflect the correlated charge and magnetic
responses. The spectra reproduce the tabulated counts to better
than $10^{-6}$ relatively. The empty sideband constrains the tail,
but its signal acceptance also depends on the compulsory decay photon.
We therefore repeat both fits with a calculable transport response.

\input{sections/05b_decay_selection}

\subsection{Other direct searches}

The XENON100 magnetic-inelastic search already used correlated recoil
and photon deposits, over 9.7--200 keV nuclear-recoil energies with
$224.6$ days and a 48-kg fiducial mass \cite{XENON100MiDM2017}.
Its published limits fix a different relation between splitting and
magnetic moment. At B0/B1, perfect-efficiency integrals in that recoil
window give $\BZeroXenonWindow$ and $\BOneXenonWindow$ events. A
dedicated response is needed for the shorter-lived present benchmarks.
The low-energy XENONnT and PandaX searches likewise cover only part of
the favored recoil interval \cite{XENONnT2025,PandaX4T2024}.
Two-dipole elastic scattering adds UV-sensitive scalar and axial
terms, a quantitative low-energy limit requires their matching
\cite{WeinerYavin2012Rayleigh,WeinerYavin2012UV}.

Heavy targets offer a complementary threshold test. At B1,
$(v_*,E_R^*)$ changes from $(785\,\mathrm{km/s},255\,\keV)$ on
xenon to approximately $(691\,\mathrm{km/s},235\,\keV)$ on tungsten.
CRESST high-energy recasts and the iodine exposure in PICO motivate
such comparisons \cite{CRESST2016,BramanteEtAl2016,SuYangYang2026,
PICO2016CF3I,PICO2023Inelastic}. Applying them to Eq.~\eqref{eq:nr-matching}
requires the full magnetic response and the calibrated high-energy
acceptance. PICO's photon-mediated analysis provides an elastic
reference \cite{PICO2022Photon}. Earth-induced excitation followed by
photon detection is another established transition-dipole probe
\cite{EbyFoxKribs2023}. And the sub-metre flights here restrict production
to material close to the instrument.

The dominant astrophysical uncertainty is the sparsely populated
high-speed tail. Annual modulation can be large and non-sinusoidal
\cite{McCabe2026}. The event's June date alone carries little evidence
for it. The LMC variation in Appendix~\ref{app:lmc} shifts the fitted
splitting substantially. At the associated momentum transfers,
$q\simeq0.25$--$0.4\,\GeV$, odd-xenon structure and two-body currents
also deserve further study \cite{KlosEtAl2013,VietzeEtAl2015}.

%% file: sections/05b_decay_selection.tex
\subsection{Decay topology and escape-selected fits}
\label{sec:decay-selection}

Energy conservation determines the outgoing speed,
\begin{equation}
 v_2^2=v^2-\frac{2(E_R+\dm)}{\mchi},\qquad \lambda_2=v_2\tau_2.
 \label{eq:outgoing-speed-selection}
\end{equation}
For B0/B1, rate-weighting the candidate's incident speeds and xenon
isotopes gives mean flights $\BZeroMeanFlight$ and $\BOneMeanFlight$ m.
The photon then travels only $\BZeroPhotonPath$ and $\BOnePhotonPath$ cm
on average before its first interaction in liquid xenon, using NIST
attenuation coefficients at density $2.9\,\mathrm{g\,cm^{-3}}$
\cite{NISTXenonAttenuation}. The metre-scale parent flight and
centimetre-scale photon transport are separate parts of the signal.

An internal decay can produce a displaced electronic recoil near
$\dm$, with additional scintillation (S1) and ionization (S2).
For deposit heights $z_1,z_2$ and electron drift speed $v_d$,
\begin{equation}
 \Delta t_{S1}\simeq t_2,\qquad
 \Delta t_{S2}\simeq t_2+(z_1-z_2)/v_d.
 \label{eq:two-pulse-timing}
\end{equation}
Thus the second S2 can arrive first. Resolved additional pulses alter
the single-scatter selection \cite{LZSelection2025}, merging, charge
loss, and deposits outside the active volume create other categories.
Their acceptance depends on both separation and delay. Delayed
recoil--photon searches have been developed specifically for this
interaction \cite{LinFinkbeiner2011,XENON100MiDM2017}.

\begin{figure*}[t]
 \includegraphics[width=\textwidth]{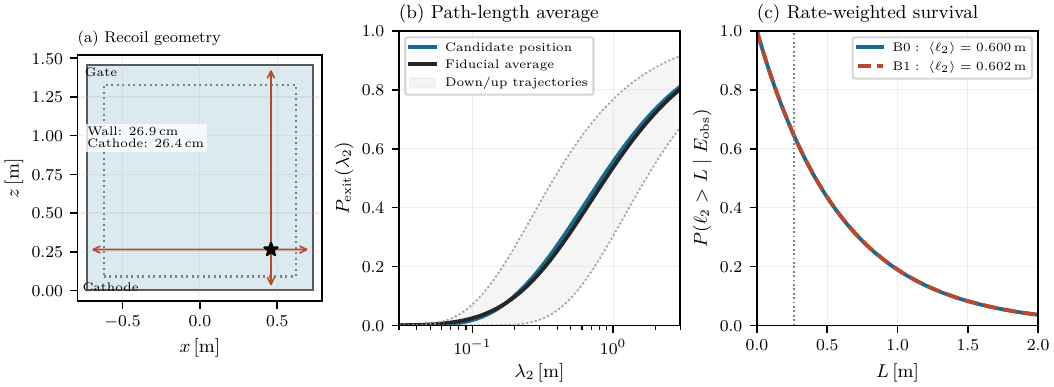}
 \caption{Transport at B0/B1. (a) Active and fiducial cylinders,
 candidate position, and example exit paths. (b) Exit probability
 at the candidate and averaged over the fiducial volume, shading
 compares upward and downward directions. (c) Flight-distance
 survival averaged over the candidate's recoil-weighted speed and
 isotope distribution. The dotted line marks the cathode stand-off.
 The curves describe the excited state before photon transport. \AIFigureNote}
 \label{fig:decay-geometry}
\end{figure*}

We approximate the active xenon as a cylinder with radius $R=0.728$ m
and height $H=1.456$ m \cite{LZGeometry2020}. The candidate's published
wall and cathode distances give $(r,z)=(0.459,0.264)$ m
\cite{LZ2026HighER}. If $D(\bm x,\hat{\bm n}_2)$ is the distance
to the first boundary, the excited-state exit probability is
\begin{equation}
 P_{\rm exit}=\exp[-D/(v_2\tau_2)].
 \label{eq:geometry-exit}
\end{equation}
Its upward, downward, outward, and inward paths are 1.192, 0.264,
0.269, and 1.187 m. Figure~\ref{fig:decay-geometry} shows how this
geometry enters the survival probability.

Our response accepts excited states that exit the TPC and rejects
internal decays. We average outgoing directions isotropically and
retain the recoil-weighted speeds. Counts use a normalized fiducial
proxy, $r<R-0.107$ m and $0.09<z<H-0.128$ m, based on the published
mean stand-offs \cite{LZ2026HighER}, the candidate intensity uses its
observed position. Denoting these weights by $a_{\rm vol}$ and
$a_{\rm evt}$ gives
\begin{align}
 N_{\rm ROI}^{\rm exit}&=\mathcal E\int\dd E_R\,
 R'_{\rm th}\epsilon_{\rm LZ}(\bar E)a_{\rm vol},\nonumber\\
 I_{\rm evt}^{\rm exit}&=\mathcal E\int\dd E_R\,
 R'_{\rm th}\epsilon_{\rm LZ}(\bar E)G_{\rm ROI}(248|\bar E)a_{\rm evt},
 \label{eq:transported-intensities}
\end{align}
where $R'_{\rm th}=\dd R_{\rm th}/\dd E_R$. The sideband uses
$a_{\rm vol}$ with its own energy response. Substituting these
quantities into Eq.~\eqref{eq:thermal-lz-likelihood} defines the
escape-selected likelihood.

\begin{figure*}[t]
 \includegraphics[width=\textwidth]{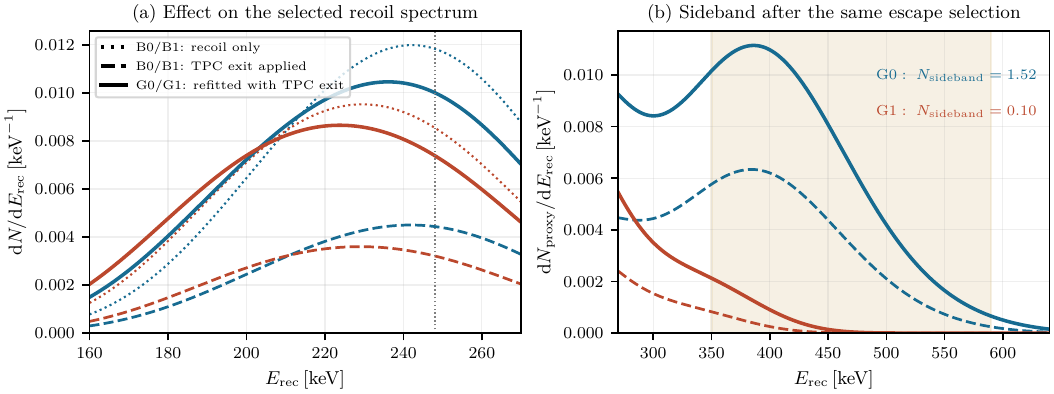}
 \caption{Selection and refit. Left: dotted curves are the recoil-only
 B0/B1 spectra, dashed curves apply TPC-exit selection at those
 parameters, and solid curves show the refitted G0/G1 spectra.
 Right: the corresponding gap and sideband predictions. Blue uses
 the candidate alone, and orange includes the empty sideband. All curves
 retain the thermal moment and common exposure. \AIFigureNote}
 \label{fig:decay-spectra}
\end{figure*}

At unchanged B0/B1 parameters the fiducial acceptances are
$\BZeroVolumeEscape$ and $\BOneVolumeEscape$, reducing the selected means
to $\BZeroExitROI$ and $\BOneExitROI$. With the moment fixed by the abundance,
refitting moves the splitting to $\GZeroDelta$ and $\GOneDelta$ keV at
G0/G1 and restores selected means $\GZeroROI$ and $\GOneROI$
(Table~\ref{tab:direct-benchmarks}). The lower threshold admits more
incident particles. Figure~\ref{fig:decay-spectra} separates the
acceptance change from this refit and shows its effect on the sideband.

TPC exit approximates one part of LZ's selection. Delayed veto windows
extend to $600\,\mu$s, with thresholds of 300 keV in the Skin and
200 keV in the outer detector \cite{LZ2026HighER}. The photons can
trigger these channels, return to active xenon after an external
decay, or escape after an internal decay. Appendix~\ref{app:decay-selection}
quantifies first-interaction transport and fiducial variations.
As a simple external-veto test, adding a fully vetoed 0.20-m path
changes the fitted splittings to $\GZeroBufferDelta$ and $\GOneBufferDelta$
keV. The full response also needs the wind-dependent outgoing
directions, event azimuth, true fiducial map, and pulse reconstruction.
The G points are benchmarks of the stated geometry approximation.

\subsection{Recoil--photon tests with further data}

Archived waveforms can be searched for an electronic recoil near
$\GOneDelta$--$\GZeroDelta$ keV following a high-energy nuclear recoil
after about a microsecond. A joint analysis can retain internal pairs,
scintillation-only deposits, veto tags, and isolated recoils
\cite{LinFinkbeiner2011,XENON100MiDM2017}. The photon energy measures
$\dm$. A delay fit including the finite-volume acceptance measures
$\tau_2$ and hence
\begin{equation}
 \mug=\left(\frac{\pi\hbar}{\tau_2\dm^3}\right)^{1/2}.
 \label{eq:followup-moment}
\end{equation}
Displacement divided by delay estimates $v_2$, up to photon transport,
and adds a kinematic mass constraint. This lifetime measurement tests
the thermal relation independently of the local density.

For orientation, the G0/G1 geometry predicts internal-decay to
isolated-recoil ratios $\GZeroPairRatio$ and $\GOnePairRatio$. With unchanged
halo and recoil efficiency, 10 tonne-years would give
$\GZeroIsolatedTen$ and $\GOneIsolatedTen$ isolated events. A pair efficiency
of 0.5 would add $\GZeroPairsTen$ and $\GOnePairsTen$ pairs. These conditional
yields set the scale for an experimental sensitivity study, backgrounds
and category migration enter its likelihood. The pair fraction cancels
exposure and local density. Its dependence on recoil energy and distance
to the wall can be compared across LZ, XENONnT, and PandaX geometries
\cite{XENONnT2025,PandaX4T2024}.

%% file: sections/05_indirect_cosmology.tex
\section{Indirect and cosmological tests}
\label{sec:indirect}

Radiative decay removes the primordial excited population. Today's
annihilation therefore proceeds predominantly from $\chi_1\chi_1$,
whereas freeze-out also involved $\chi_1\chi_2$. This population
change separates the leading relic and late-time channels
\cite{WeinerYavin2012Rayleigh,ChatterjeeLaha2022}.

\subsection{Gamma-ray lines}

Two transition-dipole insertions give, for $\mchi\gg m_Z$,
\cite{WeinerYavin2012Rayleigh,ArinaEtAl2020}
\begin{align}
 \sv_{\gamma\gamma}&=\frac{\mug^4\mchi^2}{4\pi},\nonumber\\
 \sv_{\gamma Z}&=2\tan^2\theta_W\sv_{\gamma\gamma},\nonumber\\
 \sv_{ZZ}&=\tan^4\theta_W\sv_{\gamma\gamma}.
 \label{eq:late-neutral-channels}
\end{align}
These are tree-level amplitudes in the point-dipole theory. A loop
origin for each dipole introduces additional short-distance
contributions, discussed in Sec.~\ref{sec:uv}.
The photon-weighted unresolved line is
\begin{equation}
 \sv_{\rm line}\equiv\sv_{\gamma\gamma}+\tfrac12\sv_{\gamma Z}
 =\frac{\mug^4\mchi^2}{4\pi c_W^2}.
 \label{eq:line-rate}
\end{equation}
Its photon energies are $\mchi$ and
$E_{\gamma Z}=\mchi[1-m_Z^2/(4\mchi^2)]$.
The numerical rates retain the finite-mass expressions of
Ref.~\cite{WeinerYavin2012Rayleigh}. Their high-mass branching
fractions are $(0.591,0.356,0.053)$ for $(\gamma\gamma,\gamma Z,ZZ)$
at $s_W^2=0.231$ \cite{PDG2024}.

\begin{figure*}[t]
 \includegraphics[width=\textwidth]{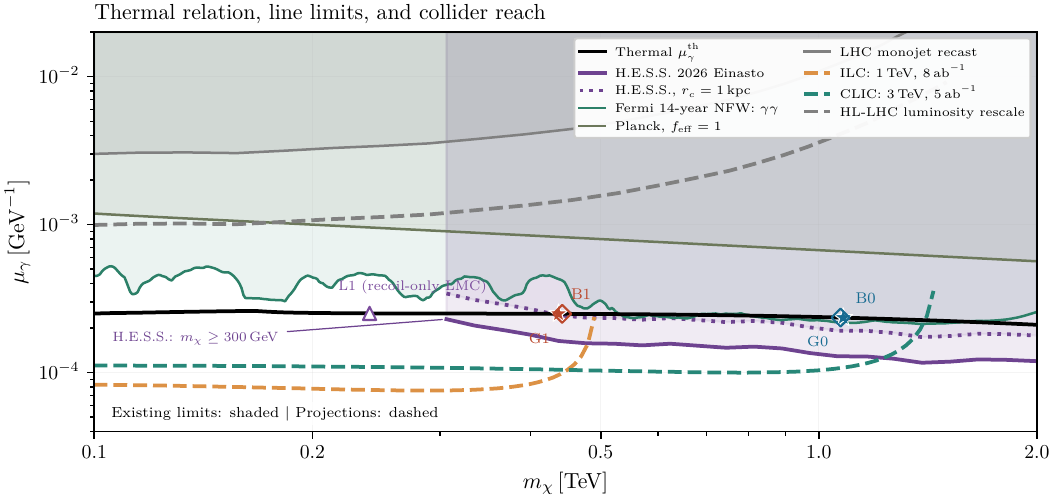}
 \caption{Constraints in the mass--moment plane. The thermal curve,
 B0/B1 stars, G0/G1 diamonds, and LMC point L1 are compared with
 H.E.S.S. Einasto and an approximate 1-kpc-core alternative
 \cite{HESS2026Lines}, the 14-year Fermi NFW $\gamma\gamma$ limit
 \cite{FosterLines2023}, and a maximal-deposition CMB envelope
 \cite{Planck2018,Slatyer2015}. Shading lies above conditional
 existing limits. Dashed collider projections use
 Ref.~\cite{ZhangEtAl2025Collider}. The HL-LHC curve rescales the
 CMS-based monojet recast by luminosity alone
 \cite{CMSMonojet2018,ArinaEtAl2020}. All Galactic comparisons use
 $\rho_\odot=0.30\,\GeV\,\mathrm{cm}^{-3}$, retaining their distinct
 spatial profiles. H.E.S.S. begins near 300 GeV. The thermal and
 applicable collider curves continue to 100 GeV. Every boundary
 assumes the corresponding point-operator hard amplitude. \AIFigureNote}
 \label{fig:combined-constraints}
\end{figure*}

Figure~\ref{fig:combined-constraints} translates published line limits
into a common moment convention, using
$\mug^{\lim}\propto(\sv^{\lim}/\mchi^2)^{1/4}$.
For H.E.S.S. we extract the observed 2026 Inner Galaxy curve
\cite{HESS2026Lines} down to its first recoverable vertex near
306 GeV. The experimental search starts at 300 GeV, so B1 lies
within its domain. Rescaling the native Einasto density from
0.39 to $0.30\,\GeV\,\mathrm{cm}^{-3}$ weakens the rate limit by
$(0.39/0.30)^2$. B0/B1 exceeds it by
$\BZeroHess$ and $\BOneHess$, respectively, the escape-selected
G0/G1 ratios are $\GZeroHess$ and $\GOneHess$. This is substantial tension for the thermal
point-dipole interpretation under Einasto.

The published cored-profile comparison gives an approximate
common-density weakening of $6.5(0.34/0.39)^2=4.94$
\cite{HESS2026Lines}. It leaves the G1 ratio near 1.1, where profile
and hard-amplitude uncertainties matter. A cored halo is therefore
a model-dependent possibility, with its viability still to be tested.

The independent 14-year Fermi analysis extends from 10 GeV to 2 TeV
\cite{FosterLines2023}. We retain its NFW morphology and rescale
the density from 0.4 to $0.30\,\GeV\,\mathrm{cm}^{-3}$.
Figure~\ref{fig:combined-constraints} uses the $\gamma\gamma$
component alone. Its ratios to the Fermi limit are
$\BOneFermiGG$ at B1 and $\GOneFermiGG$ at G1, combining unresolved
lines gives $\BOneFermiBoth$ and $\GOneFermiBoth$, respectively.
The observed Fermi curve weakens locally near these masses.
At G0 the $\gamma\gamma$ ratio is $\GZeroFermiGG$,
with H.E.S.S. still the stronger test.
The lower-mass LMC point L1 has ratio $\LOneFermiGG$ and lies below
the H.E.S.S. range. Its recoil-only acceptance remains to be
replaced by a joint decay response (Appendix~\ref{app:lmc}).

Fermi's Pass-8 line search and MAGIC provide independent Galactic
data \cite{FermiLATLines2015,MAGICLines2023}, HAWC dwarf searches
extend the high-mass comparison beyond the inner Milky Way
\cite{HAWCLines2020}. Correlated-line methods are developed in
Ref.~\cite{DeLaTorreLuque2024}. CTA projections indicate substantial
multi-TeV reach, although their quoted gain uses an older H.E.S.S.
reference \cite{CTA2024Lines}. At fixed profile, improving a rate
limit by a factor of two lowers the moment boundary by $2^{-1/4}$.

\subsection{Continuum, energy deposition, and solar capture}

The minimal continuum comes from $Z$ decays in $\gamma Z$ and $ZZ$.
The equivalent two-$Z$ rate, $\sv_{\gamma Z}/2+\sv_{ZZ}$,
is below representative continuum sensitivities near the fitted
masses \cite{FermiLATDwarfs2023,HESS2022Continuum,CaloreEtAl2022}.
A mixed-channel comparison folds the channel yields
\cite{CirelliEtAl2011PPPC} into the photon or antiproton likelihood.
Extra UV channels can make Fermi dwarfs, H.E.S.S./CTA continuum data,
and AMS-02 antiprotons more restrictive
\cite{CTAConsortium2021,AMS2016Antiprotons,AMS2025Antiprotons}.
The broad $Z$-decay neutrino spectrum similarly needs its own
response convolution, a neutrino-line limit probes a different
spectral shape \cite{IceCubeLines2023}.

Planck constrains
\begin{equation}
 p_{\rm ann}=f_{\rm eff}\frac{\sv}{\mchi}
 <3.2\times10^{-28}\,\mathrm{cm^3\,s^{-1}\,GeV^{-1}}
 \label{eq:planck-limit}
\end{equation}
at 95\% confidence \cite{Planck2018,Slatyer2015}.
Even $f_{\rm eff}=1$ gives $\BZeroCMB\%$ and $\BOneCMB\%$ of this
bound at B0 and B1. Their short excited-state lifetimes also precede
big-bang nucleosynthesis,
sub-MeV de-excitation photons give
negligible photodissociation constraints
\cite{ForestellEtAl2018,CarrilloGonzalezToro2021}.
A separate abundance-setting sector brings its own energy-deposition
and decay constraints.

\label{sec:solar}
Solar infall opens scattering on heavy elements
\cite{Gould1987,NussinovWangYavin2009,MenonEtAl2010}.
Our first-capture calculation gives
$C_\odot=(\BZeroSolarCapture,\BOneSolarCapture)\,\mathrm{s}^{-1}$
at B0 and B1, respectively, dominated by Fe and Ni. The charge response survives on
their spin-zero isotopes even when odd xenon dominates LZ.
Appendix~\ref{app:solar} gives the integral, isotope content, and
dependence on the captured distribution.
Recent solar tests of LZ interpretations
\cite{PospelovRamani2026,DiMauroShaikh2026Solar,BoseEtAl2026Solar}
underline the importance of this channel. For a transition dipole,
both the orbital distribution and the $\gamma Z,ZZ$ neutrino yields
enter an IceCube or Super-Kamiokande comparison
\cite{CatenaHellstrom2018,BlennowEtAl2015,IceCubeSolar2025}.
We give first-capture and equilibration estimates, a solar-neutrino
exclusion requires further evolution and detector calculation.

%% file: sections/06_uv_and_future.tex
\section{Possible UV completions and Collider tests}
\label{sec:uv}

\subsection{UV completion with weak messengers and beyond}
\label{sec:weak-messengers}

The renormalizable construction of Ref.~\cite{WeinerYavin2012UV}
contains a singlet Dirac precursor $\Psi$, a vectorlike fermion
$\psi$, and a complex scalar $\phi$, with both messengers in
$(\bm2,Y=1/2)$. Its relevant terms are
\begin{align}
 \Lagr\supset{}&
 -\mchi\bar\Psi\Psi-\frac{\dm}{4}
   (\bar\Psi^c\Psi+\bar\Psi\Psi^c)
 -M_f\bar\psi\psi-M_s^2\phi^\dagger\phi\nonumber\\
 &+(\lambda\bar\psi\Psi\phi+\mathrm{h.c.}).
 \label{eq:wy-lagrangian}
\end{align}
The Majorana term gives masses $\mchi\mp\dm/2$ at this order.
For example, assigning $\Psi,\psi$ odd parity and $\phi$ even
allows small decay portals
\begin{equation}
 \Delta\Lagr=\epsilon_\psi\bar\psi_iH^i\Psi
 -m_{\phi H}^2\phi_i^\dagger H^i+\mathrm{h.c.}
 \label{eq:messenger-decay-portals}
\end{equation}
Their size controls messenger lifetimes and induced Higgs or weak
scattering. The dipole-dominated hierarchy requires these additional
recoil amplitudes to remain subleading.

For $N$ identical copies in a representation of dimension $d_R$ and
hypercharge $Y$, define the charge sum $Q=d_RY$. In the
heavy-messenger limit \cite{WeinerYavin2012UV},
\begin{align}
 \mug&=\frac{NQe\lambda^2}{32\pi^2M_f}F_2(r),\qquad r=M_f/M_s,\nonumber\\
 F_2(r)&=\frac{2r^2(r^2-1-\ln r^2)}{(1-r^2)^2},\quad F_2(1)=1.
 \label{eq:messenger-moment}
\end{align}
Thus $\gM=(NQ\alpha_\lambda/2\pi)(\mchi/M_f)F_2$,
where $\alpha_\lambda=\lambda^2/(4\pi)$.
For equal messenger masses $M$, the exact low-momentum factor is
\begin{equation}
 F_2(0;a)=2\int_0^1\dd z\,(1-z)
 \frac{1+2az(1-z)}{1-a^2z(1-z)},
 \label{eq:messenger-exact-q0}
\end{equation}
with $a=\mchi/M$, from Ref.~\cite[Appendix~A]{WeinerYavin2012UV}.
The required product is
\begin{equation}
 NQ\alpha_\lambda=\frac{8\pi M\mug}{eF_2(0;\mchi/M)}.
 \label{eq:fitted-messenger-requirement}
\end{equation}
At the illustrative scale $M=1.14$ TeV, B0/B1 require
$\BZeroMessenger$ and $\BOneMessenger$, respectively. This exceeds the one-copy,
unit-charge perturbative realization. Large charge or multiplicity
also affects gauge running and direct production. Near-degenerate
messengers contribute to freeze-out and require an extended
coannihilation calculation \cite{GriestSeckel1991}.

The same charged loop produces scalar and pseudoscalar Rayleigh
operators, schematically $\bar\chi\chi B_{\mu\nu}B^{\mu\nu}$
and $i\bar\chi\gamma^5\chi B_{\mu\nu}\widetilde B^{\mu\nu}$, with
weak-field counterparts \cite{WeinerYavin2012Rayleigh,WeinerYavin2012UV}.
For $a=Y,2$, their heavy-messenger coefficients scale as
\begin{equation}
 C_{Ra}=\frac{Ng_a^2\lambda^2 C_a}{48\pi^2M_f^3}F(r),
 \quad
 \widetilde C_{Ra}=\frac{Ng_a^2\lambda^2 C_a}{48\pi^2M_f^3}
 \widetilde F(r).
 \label{eq:rayleigh-matching}
\end{equation}
Here $C_Y=d_RY^2$ and $C_2=T(R)$, with both equal to $1/2$
for the doublet. A dipole scales as $NQ$ and a Rayleigh amplitude
as $NC_a$. Their different charge dependence permits distinct
line spectra at a common measured moment.
In the Rayleigh-dominated doublet example of
Ref.~\cite{WeinerYavin2012UV},
$\sigma_{\gamma\gamma}/(\sigma_{\gamma Z}/2)\simeq2.2$ and
$\sigma_{\rm total}/\sigma_{\rm line}\simeq5$, compared with
3.33 and 1.30 for the pure hypercharge dipole.
The complete hard amplitude and abundance must be matched together. 

Besides the weak realizations discussed above, several strong dynamics have been studied to produce the dipole transitions; they are proposed to better explain the order unity $g_M$. We refer readers to Ref. \cite{ArandaBarajasCembranos2016, AsadiEtAl2026Magnetic} for details.

\subsection{Collider reach}
\label{sec:collider}

Monojet and electroweak production studies constrain the timelike
dipole \cite{FortinTait2012,BargerEtAl2012,ArinaEtAl2020},
lepton-collider studies give complementary projections
\cite{ZhangEtAl2022Collider,ZhangEtAl2025Collider}.
In the nearly degenerate limit, inclusive $\chi_1\chi_2$ production
has the Dirac pair normalization. At the fitted splittings,
$c\tau_2$ is hundreds of metres and the secondary photon is soft,
so an invisible-pair recast is appropriate under the point-production
assumption.

Figure~\ref{fig:combined-constraints} uses the magnetic curves in
Ref.~\cite[Fig.~10]{ZhangEtAl2025Collider}, converted from Bohr
magnetons to $\GeV^{-1}$. Its 1-TeV ILC and 3-TeV CLIC projections
use 8 and $5\,\mathrm{ab}^{-1}$ of polarized running, respectively.
At B1 the thermal moment is $\BOneILC$ times the ILC reach and
$\BOneCLIC$ times the CLIC reach. B0 is beyond the ILC pair
threshold and is $\BZeroCLIC$ times the CLIC reach.
These comparisons inherit the study's background assumptions.

The hadron-collider guide derives from the CMS
$35.9\,\mathrm{fb}^{-1}$ monojet recast
\cite{CMSMonojet2018,ArinaEtAl2020}.
For fixed cuts and background-dominated statistical errors,
$\mug^{\lim}\propto L^{-1/4}$, extrapolating to
$3\,\mathrm{ab}^{-1}$ improves the moment reach by 0.331.
Even this projection remains above the B0/B1 moments by factors
$\BZeroHLDistance$ and $\BOneHLDistance$, respectively. It holds energy, acceptance,
and background shapes fixed, systematics weaken the gain.

Resolved-messenger production and decays have been studied in
Refs.~\cite{LiuShuveWeinerYavin2013,PrimulandoSalvioniTsai2015}.
For detector-stable unit-charge Drell--Yan fermions, CMS reaches
about 1.14 TeV, its degenerate-stau model reaches 0.69 TeV
\cite{CMSHSCP2025}. These topology-specific limits motivate the
matching example above. Prompt leptonic decays or compressed
charged states connect to multilepton and disappearing-track
searches \cite{ATLASThreeLeptons2021,ATLASDisappearing2022}.
The applicable limits depend on the spectrum, branching fractions,
and lifetimes.

\subsection{Connecting low- and high-energy measurements}

A recoil--photon measurement determines $\dm$ and $\mug$ through
Eq.~\eqref{eq:followup-moment}. At the G1 moment, the H.E.S.S.
Einasto limit requires the line amplitude to be at most
$1/\sqrt{\GOneHess}\simeq\GOneAmplitudeLimit$ of its point-dipole
value. This is a quantitative target for hard matching at a fixed
low-energy moment. Independent Rayleigh amplitudes can interfere
differently in each final state \cite{WeinerYavin2012UV}.

At the thermal benchmark masses the two lines are separated by
$10^{-3}$--$10^{-2}$ fractionally, below the resolution of
atmospheric Cherenkov telescopes \cite{HESS2026Lines,CTA2024Lines}.
The unresolved-line to continuum ratio therefore offers a practical
test of the messenger representation. Collider production supplies
another hard-momentum measurement, while solar neutrinos probe both
annihilation and any UV-induced elastic cooling.
These tests complement the common low-energy transition measured
by isolated recoils and delayed pairs. Appendix~\ref{app:secluded}
shows how the LZ fit changes when an additional channel allows a
smaller physical moment.

%% file: sections/07_conclusions.tex
\section{Conclusions}
\label{sec:conclusions}

A transition magnetic dipole connects a high-energy nuclear recoil
to a delayed photon. In the minimal hypercharge theory, the thermal
abundance fixes a moment near $2.4\times10^{-4}\,\GeV^{-1}$ at the
LZ-preferred masses. Fitting the recoil then determines a conditional
mass--splitting band. The empty sideband favors a lower mass and
a softer high-energy tail.

The predicted excited state travels about 0.6 m, so its decay affects
the isolated-recoil selection. A position-aware TPC-exit response
gives G0/G1 at $(\GZeroMass\,\TeV,\GZeroDelta\,\keV)$ and
$(\GOneMass\,\TeV,\GOneDelta\,\keV)$, with
$\GZeroSideband$ and $\GOneSideband$ sideband-proxy events, respectively.
The corresponding H.E.S.S. Einasto ratios, about 11 and 5.5,
disfavor these thermal point-dipole benchmarks.
The cored-profile comparison and LMC variation illustrate
astrophysical sensitivity, their full recoil--photon acceptance
remains important.

A free moment below the thermal ceiling broadens the LZ support
over 0.1--100 TeV. The global maxima remain near G0/G1, while a
6.8-TeV profiled solution lies below the quoted Einasto line bound.
Its abundance requires an additional annihilation channel. Charged-messenger
matching further shows that the thermal moments demand a large
multiplicity--coupling product, while independent hard amplitudes
change the line and collider predictions.

A joint analysis of isolated recoils, delayed recoil--photon pairs,
and veto tags can test the transition directly. Photon energies and
delays would measure the splitting and moment, the relative event
populations would constrain the decay geometry. Comparisons with
line searches and a specified messenger spectrum would then test
the thermal interpretation.

%% file: sections/note_added.tex
\section*{Note added}
\label{sec:note-added}

During the completion of this work, several related studies appeared.
Asadi, Batz, Fox, Homiller, and Kribs~\cite{AsadiEtAl2026Magnetic}
proposed a magnetic dark-baryon interpretation of LZ230616. Their
model shares photon-mediated inelastic scattering and correlated
de-excitation, with a confining origin for the states and a fixed
dimensionless moment. The present study determines the moment
from the minimal hypercharge relic abundance and includes an
explicit escape response in the recoil fit. Appendix~\ref{app:secluded}
compares their preferred points with our free-moment extension.
Two other recent applications use magnetic interactions in different
ways: Ref.~\cite{LiangEtAl2026Magnetic} considers boosted elastic
scattering through an isoscalar contact tensor interaction, while
Ref.~\cite{ZhuEtAl2026DarkPhoton} introduces a weak transition dipole
to deplete excited states in a dark-photon portal.

%% file: sections/acknowledgments.tex
\begin{acknowledgments}
We thank Christina Gao, Jinhui Guo, Tae Kim, Aidi Yang, and Yi-Ming Zhong for useful discussions.
This work was supported by the GRF Grants No. 11302824 and No. 11310925 from the Research Grants Council, University Grants Committee, and the Grants No. 9610645 and No. 7020130 from the City University of Hong Kong.
OpenAI's ChatGPT and Codex assisted with
literature searches, derivations
and consistency checks, development of numerical code and figures, and
drafting and editing the manuscript. The authors take responsibility
for the calculations, references, and scientific conclusions.
\end{acknowledgments}

%% file: sections/appendix_lz_sideband.tex
\section{Recoil response and high-energy sideband}
\label{app:lz-sideband}

The public search uses $3<S1c<600$ phd,
$10^{2.75}<S2c<10^{4.15}$ phd, and $S2>645$ phd.
The disjoint high-energy sideband has $800<S1c<1700$ phd and
$10^{2.75}<S2c<10^{4.3}$ phd
\cite{LZ2026HighER,RoddEtAl2026Sideband}.
The intervening $S1c$ interval has no public WIMP-like event count.
Table~\ref{tab:lz-sideband-bookkeeping} gives the published control
and science cells. The 5.4-tonne annulus is the volume between the
nested 4.7- and 5.4-tonne boundaries, with a separate acceptance.

\begin{table}[tb]
\caption{High-energy-sideband entries from LZ Supplemental Table S6
\cite{LZ2026HighER}. Each cell gives simulated/observed counts.
Simulation entries describe only the multiple-scintillation,
single-ionization background component.}
\label{tab:lz-sideband-bookkeeping}
\begin{ruledtabular}
\begin{tabular}{lcc}
Cell & Wall & Below cathode\\
\midrule
4.7-t science & 0.1 / 0 & 0.5 / 0\\
4.7-t prompt veto & 0.3 / 1 & 0.5 / 0\\
5.4-t annulus science & 0.5 / 0 & 21.5 / 18\\
5.4-t annulus prompt & 2.2 / 0 & 5.2 / 3\\
\end{tabular}
\end{ruledtabular}
\end{table}

Let $G(E|\bar E,\sigma)$ be a normalized Gaussian with the scale and
width of Eq.~\eqref{eq:profile-energy-response}, and define
$P_{ab}(\bar E)=\Phi[(b-\bar E)/\sigma]-\Phi[(a-\bar E)/\sigma]$.
Our conditional search-window kernel is
\begin{equation}
 G_{\rm ROI}(E|\bar E)=
 \frac{G(E|\bar E,\sigma)}{P_{5.4,269.9}(\bar E)}
 \mathbf 1_{[5.4,269.9]}(E).
 \label{eq:conditional-roi-kernel}
\end{equation}
All energy bounds are in keV. This normalization preserves the
selected yield already encoded in LZ's efficiency. The efficiency
is interpolated through a digitization of LZ Fig.~S2 with its
published 50\% endpoints imposed \cite{LZ2026HighER}.

The gap and sideband predictions use
\begin{align}
 N_{\rm gap}&=\mathcal E\int\dd E_R\,
 \frac{\dd R_{\rm th}}{\dd E_R}P_{269.9,350}(\bar E),\nonumber\\
 N_{\rm sideband}&=\mathcal E\int\dd E_R\,
 \frac{\dd R_{\rm th}}{\dd E_R}P_{350,590}(\bar E).
 \label{eq:profile-sideband-integrals}
\end{align}
These are unit-response energy proxies. The escape-selected version
inserts $a_{\rm vol}$ into both integrals, all quantities retain the
same thermal moment, annual spectrum, and fitted scale.
For B0/B1 the gap predictions are $\BZeroGap$ and $\BOneGap$.
The gap contributes no likelihood term.

At zero observed sideband events, a fixed background $b$ contributes
$-b$ to $\ln\mathcal L$. It cancels from parameter likelihood ratios
but enters the probability $P(0)=\exp[-N_{\rm sideband}-b]$.
The familiar zero-background 90\% Poisson scale is 2.30 events,
subtracting a fixed 0.6 gives 1.70, while $CL_s$ returns 2.30
\cite{PDG2024,Read2002}. These are counting comparisons conditional
on the response and background prescription.

\begin{table}[tb]
\caption{Response variations of the recoil-only thermal fit.
Masses are in TeV and splittings in keV. Unlisted settings equal
the B0/B1 reference. The sideband multiplier is a sensitivity
choice, with no probability distribution assigned.}
\label{tab:response-fits}
\begin{ruledtabular}
\begin{tabular}{lcc}
Setting & $\mchi$ & $\dm$\\
\midrule
Without sideband (B0) & \BZeroMass & \BZeroDelta\\
With sideband (B1) & \BOneMass & \BOneDelta\\
\input{sections/response_table_rows.tex}
\end{tabular}
\end{ruledtabular}
\end{table}

Table~\ref{tab:response-fits} separates the scale and sideband
choices. The upper-edge extension to 680 keV has little effect
because B1's high-energy population is already small. The lower
edge and sideband normalization matter more.
For context, the Higgsino study of
Ref.~\cite{RoddEtAl2026Sideband} found about 4.9 or 3.7 sideband
events per search-window event for Helm or Vietze responses,
respectively. A matched comparison also requires the same
efficiency and energy kernel.

A full prediction replaces these proxies by
\begin{equation}
 N_{\rm sideband}=\mathcal E\int\dd E_R\,\frac{\dd R}{\dd E_R}
 \mathcal K_{\rm sideband}(E_R,\bm x,\hat{\bm p}_2,\lambda_2).
 \label{eq:physical-sideband-count}
\end{equation}
The kernel includes decay, photon interactions, S1/S2 formation,
vetoes, and migration between categories. LZ230616 was 26.4 cm above
the cathode and 26.9 cm inside the true wall
\cite{LZ2026HighER}. These distances motivate the transport
calculation in Sec.~\ref{sec:decay-selection}. Signal injections
with the actual live-time and detector maps would replace the
uniform annual and cylindrical averages used here.

%% file: sections/response_table_rows.tex
Without sideband, fixed scale & 1.101 & 355.9 \\
With sideband, fixed scale & 0.452 & 328.7 \\
$A_{\rm sideband}=0.25$ & 0.533 & 336.1 \\
$A_{\rm sideband}=0.5$ & 0.478 & 331.1 \\
$A_{\rm sideband}=2$ & 0.407 & 322.7 \\
Sideband upper edge 680 keV & 0.437 & 326.7 \\
Sideband lower edge 375 keV & 0.462 & 329.5 

%% file: sections/appendix_nreft.tex
\section{Nonrelativistic matching}
\label{app:nreft}

We use metric $(+,-,-,-)$ and momentum transfer
$q=p_2-p_1=k_1-k_2$, where $p_i$ and $k_i$ are dark-matter and
nucleon momenta. Write $q=|\bm q|$ in nonrelativistic expressions.
With $F_1^N(0)=Q_N$ and $F_2^N(0)=\kappa_N=g_N/2-Q_N$, photon
exchange gives, up to a common phase,
\begin{align}
 \mathcal M_N=\frac{e\mug}{q_{\rm Lor}^2}
 &[\bar u_2i\sigma^{\mu\nu}q_\nu u_1]\nonumber\\
 {}\times&\bar u_N'\left[Q_N\gamma_\mu
 -\frac{i\kappa_N}{2m_N}\sigma_{\mu\alpha}q^\alpha\right]u_N .
 \label{eq:matching-rel-amplitude}
\end{align}
The target Pauli current carries $-q$. Spinors obey $\bar uu=2m$,
and $\mathcal M_{\rm NR}=\mathcal M_N/(4\sqrt{m_1m_2}m_N)$.
Energy conservation fixes
$\bm v\cdot\bm q=-q^2/(2\mu_{\chi N})-\dm$, so the inelastic
transverse velocity is \cite{BarelloChangNewby2014}
\begin{equation}
 \bm v_\perp=\bm v+\frac{\bm q}{2\mu_{\chi N}}
 +\frac{\dm\bm q}{q^2},\qquad \bm v_\perp\cdot\bm q=0.
 \label{eq:vinel-perp-app}
\end{equation}
The operator basis is
\begin{align}
 O_1&=\mathbf 1,\qquad O_4=\bm S_\chi\cdot\bm S_N,\nonumber\\
 O_5&=i\bm S_\chi\cdot(\bm q/m_N\times\bm v_\perp),\nonumber\\
 O_6&=(\bm S_\chi\cdot\bm q/m_N)(\bm S_N\cdot\bm q/m_N).
 \label{eq:nr-operators}
\end{align}

Using the unequal-mass Gordon identity,
$\bar u_2i\sigma^{\mu\nu}q_\nu u_1
=\bar u_2[(m_1+m_2)\gamma^\mu-(p_1+p_2)^\mu]u_1$,
the leading Pauli reduction is
\begin{align}
 \mathcal M_{\rm NR}^N=e\mug\bigg[
 &\frac{Q_N}{2\mchi}O_1-\frac{2Q_Nm_N}{q^2}O_5\nonumber\\
 &+\frac{2(Q_N+\kappa_N)}{m_N}
 \left(O_4-\frac{m_N^2}{q^2}O_6\right)\bigg].
 \label{eq:matching-reduced}
\end{align}
This yields Eq.~\eqref{eq:nr-matching}. The longitudinal
$\dm\bm q/q^2$ term vanishes inside the cross product in $O_5$.
Reversing momentum flow reverses $O_5$, and the coefficient must use
the convention of the nuclear-response calculation
\cite{AnandFitzpatrickHaxton2013,CirelliDelNobilePanci2013}.

For a spin-$1/2$ point target, the spin trace is
\begin{equation}
 \overline{|\mathcal M_{\rm NR}|^2}=e^2\mug^2
 \left[Q_N^2\left(\frac{1}{4\mchi^2}+\frac{v_\perp^2}{q^2}\right)
 +\frac{g_N^2}{8m_N^2}\right].
 \label{eq:point-spin-trace}
\end{equation}
The factor $1/8$ is the transverse two-spin trace.
The differential cross section is
$m_T\overline{|\mathcal M_{\rm NR}|^2}/(2\pi v^2)$.
For the nuclear charge term this recovers
\cite{ChangWeinerYavin2010}
\begin{align}
 \frac{\dd\sigma_{DZ}}{\dd E_R}
 &=\frac{\alpha Z^2\mug^2}{E_R}F_E^2(q)K_{DZ},\nonumber\\
 K_{DZ}&=1-\frac{E_R}{v^2}\left(\frac{1}{2m_T}+\frac1{\mchi}\right)\nonumber\\
 &\quad-\frac{\dm}{v^2}\left(\frac1{\mu_{\chi T}}+\frac{\dm}{2m_TE_R}\right).
 \label{eq:dz-cross-section}
\end{align}
The $\dm$ terms are equivalently contained in $v_\perp^2=v^2-v_{\min}^2$.
Finite nuclei retain the full charge, orbital, and spin interference.

An independent calculation of exact free-spinor amplitudes tests
243 mass, angle, speed, and splitting combinations. The maximum
relative matrix residual is $1.67\times10^{-5}$, decreasing to
$4.64\times10^{-7}$ at $v/c=0.001$, as expected for omitted
relativistic terms. The analytic spin trace agrees to
$7\times10^{-16}$, and a hydrogen-target rate test checks the
unit conversion to $10^{-7}$ relatively. Reversing $c_5$ alone
fails the spin-amplitude test. A separate free-target comparison
reaching $q=0.4$ GeV finds corrections up to 2.8\%. The nuclear
uncertainty also involves finite-momentum many-body structure
\cite{KlosEtAl2013,VietzeEtAl2015}.

The two halo moments required by magnetic scattering are
\begin{align}
 \eta(w)&=\int_w^\infty\frac{f_1(v)}v\,\dd v,\nonumber\\
 \eta_\perp(w)&=\int_w^\infty\frac{f_1(v)}v(v^2-w^2)\,\dd v
 =2\int_w^\infty v\eta(v)\,\dd v.
 \label{eq:halo-moments-app}
\end{align}
Both are reconstructed from one normalized speed density, including
in the LMC comparison \cite{DelNobileEtAl2012Magnetic}.

%% file: sections/appendix_relic.tex
\section{Annihilation kernels and abundance calculation}
\label{app:relic}

For equilibrium fractions $r_i$, the total-density effective rate is
\begin{equation}
 \sv_{\eff}=2r_1r_2\sv_{12}+(r_1^2+r_2^2)\sv_N^{(2)}.
 \label{eq:sigmaeff-full-app}
\end{equation}
Here $r_2/r_1=(1+\dm/\mchi)^{3/2}e^{-\dm/T}$.
The two Majorana states each have two spin states
\cite{GriestSeckel1991}. In the convention $Q=T_3+y/2$, summing
Standard Model chiral fermions gives
$\sum_fN_cy_f^2=22_q+15_{\ell^\pm}+3_\nu=40$.
The zero-velocity high-mass physical pair rates are
\cite{ArinaEtAl2020}
\begin{equation}
 (\sigma_{12}v)_0=\frac{e^2\mug^2}{\pi c_W^4}
 \left(\frac{40}{32}+\frac2{64}\right),\quad
 (\sigma_N^{(2)}v)_0=\frac{\mug^4\mchi^2}{4\pi c_W^4}.
 \label{eq:sigma12-summed-app}
\end{equation}
The last one-insertion term includes $W^+W^-$ and $Zh$.
Equal populations give Eq.~\eqref{eq:moment-thermal-scaling}.

\subsection{Finite-mass one-insertion rates}

Define $\beta_\chi=\sqrt{1-4\mchi^2/s}$,
$r_f=m_f^2/s$, $\beta_f=\sqrt{1-4r_f}$, and
\begin{align}
 P_Z(s)&=\frac{s}{s-m_Z^2+im_Z\Gamma_Z},\nonumber\\
 C_V^f&=Q_f-\frac{T_3^f/2-Q_fs_W^2}{c_W^2}P_Z,\qquad
 C_A^f=-\frac{T_3^f}{2c_W^2}P_Z.
 \label{eq:finite-dipole-currents}
\end{align}
The physical fermion rate is
\begin{align}
 (\sigma v_{\rm M})_{f\bar f}
 ={}&\frac{N_c\alpha\mug^2}{3}
 \left(1+\frac{8\mchi^2}{s}\right)\beta_f\nonumber\\
 &\times[(1+2r_f)|C_V^f|^2+(1-4r_f)|C_A^f|^2],
 \label{eq:finite-fermion-rate}
\end{align}
where $v_{\rm M}=2\beta_\chi$ is the centre-of-mass M\o ller
velocity. The normalization tends to $\alpha\mug^2$ at threshold
for a massless unit-charge fermion coupled only to the photon.

Define $r_W=m_W^2/s$, $\beta_W=\sqrt{1-4r_W}$, and
$k_{Zh}^2=s\beta_{Zh}^2/4$, with
$\beta_{Zh}=\lambda^{1/2}(s,m_Z^2,m_h^2)/s$ in terms of the
K\"all\'en function. The boson rates are
\begin{align}
 (\sigma v_{\rm M})_{WW}={}&\frac{\alpha\mug^2}{48c_W^4}
 \left(1+\frac{8\mchi^2}{s}\right)|P_Z|^2\nonumber\\
 &{}\times\beta_W^3(1+20r_W+12r_W^2),\label{eq:finite-ww-rate}\\
 (\sigma v_{\rm M})_{Zh}={}&\frac{\alpha\mug^2\beta_{Zh}}{12c_W^4}
 \frac{(s+8\mchi^2)(3m_Z^2+k_{Zh}^2)}
 {(s-m_Z^2)^2+m_Z^2\Gamma_Z^2}.
 \label{eq:finite-zh-rate}
\end{align}
Each channel vanishes below its physical threshold. Combining the
photon and $Z$ diagrams before squaring preserves the longitudinal
$W$ cancellation \cite{Heo2010,ArinaEtAl2020}.
We use the tree input scheme $m_W=c_Wm_Z$, with
$m_Z=91.1876$, $\Gamma_Z=2.4952$, $m_h=125.25$, $m_t=172.69$ GeV,
and $s_W^2=0.23122$ \cite{PDG2024}.
Independent fermion traces and sums over physical $W$ polarizations
reproduce these rates.

\subsection{Thermal integration and neutral channels}

We thermally average each kernel with the invariant integral of
Ref.~\cite{GondoloGelmini1991}, using
$\sigma(s)=(\sigma v_{\rm M})/(2\beta_\chi)$.
Splitting the quadrature at final-state thresholds retains thermally
opened channels. In the massless limit the one- and two-insertion
thermal factors are $1-5/(2x)+5/x^2$ and
$1+3/(2x)+1/(2x^2)$, respectively, with $x=\mchi/T$.
These expansions include flux and thermal-measure corrections and
agree with direct integration at the per-mille level near $x=25$.

For the neutral channels we retain the finite-mass $s$ waves of
Ref.~\cite{WeinerYavin2012Rayleigh}. At $\dm=0$, the high-mass
$\gamma Z$ and $ZZ$ rates in Eq.~\eqref{eq:late-neutral-channels}
are multiplied by
\begin{align}
 F_{\gamma Z}&=(1-z_Z)(1+z_Z)^2,\nonumber\\
 F_{ZZ}&=(1-4z_Z)^{3/2}\frac{(1+2z_Z)^2}{(1-2z_Z)^2},
 \quad z_Z=\frac{m_Z^2}{4\mchi^2}.
 \label{eq:finite-mass-lines}
\end{align}
The numerical benchmarks also retain the splitting.
The neutral thermal average uses the second velocity factor above.
Replacing its entire finite-mass sum by the massless sum shifts
the abundance by at most $\ThermalNeutralMassPercent\%$ in the
100-GeV--5-TeV checks, where this channel is small.

The yield $Y=(n_1+n_2)/s$ obeys
\begin{equation}
 \frac{\dd Y}{\dd x}=-\frac{s\sv_{\eff}}{Hx}
 \left(1+\frac13\frac{\dd\ln g_s}{\dd\ln T}\right)(Y^2-Y_{\rm eq}^2).
 \label{eq:boltzmann-y}
\end{equation}
Entropy conservation places the derivative in the numerator
\cite{DreesEtAl2015}. We interpolate the published
$g_s(T),g_\rho(T)$ table \cite{SaikawaShirai2018} and integrate
from $x=5$ to $10^7$, checking the solution with independent
implicit solvers and initial points.
Off-grid thermal roots have abundance residuals below
$\ThermalInterpolationError$. Replacing the late integration tail
by direct quadrature changes the abundance by at most
$\ThermalLateTailError$.
The small splitting is retained in the equilibrium weights.
Plasma conversions, of order $\alpha\mug^2T^3$, comfortably
maintain chemical equilibrium for the thermal benchmarks
\cite{CarrilloGonzalezToro2021,ChatterjeeLaha2022}.

Numerical precision is finer than the hard-rate uncertainty.
Electroweak running, thermal masses, and finite-mass neutral velocity
corrections are perturbative improvements. Light mediators,
nearby charged states, or independent Rayleigh operators
change the annihilation model. The 100-TeV extension in
Appendix~\ref{app:secluded} is a conditional point-operator envelope
in that regime.

%% file: sections/appendix_lmc.tex
\section{An LMC-sensitive halo}
\label{app:lmc}

The Large Magellanic Cloud (LMC) modifies both the stripped-particle
population and the native Milky-Way halo
\cite{SmithOrlikEtAl2023}. Endothermic scattering is especially
sensitive to the resulting fast tail \cite{FanReece2026}.
We use the total present-day detector-frame halo integral in
Ref.~\cite[Fig.~11]{SmithOrlikEtAl2023}, retaining its boost convention
and replacing the SHM speed distribution. The local density stays
at $0.30\,\GeV\,\mathrm{cm}^{-3}$, so this comparison tests the
published speed shape. Its circular speed is 220 km/s, compared
with 238 km/s in our SHM, isolating the LMC perturbation at identical
Galactic inputs would require the simulation particles.

The extracted curve extends to $v_c=950$ km/s. We remove the
unresolved tail and renormalize:
\begin{align}
 \eta_c(v)&=\frac{\eta_{\rm pub}(v)-\eta_{\rm pub}(v_c)}{\mathcal N_c}
 \Theta(v_c-v),\nonumber\\
 \mathcal N_c&=\int_0^{v_c}
 [\eta_{\rm pub}(v)-\eta_{\rm pub}(v_c)]\,\dd v.
 \label{eq:lmc-truncation}
\end{align}
The subtraction removes above-cut particles from every threshold.
It gives a normalized positive speed density
$f_{1,c}(v)=-v\,\dd\eta_c/\dd v$, the second magnetic halo moment
follows from Eq.~\eqref{eq:halo-moments-app}.

\begin{figure*}[t]
 \includegraphics[width=\textwidth]{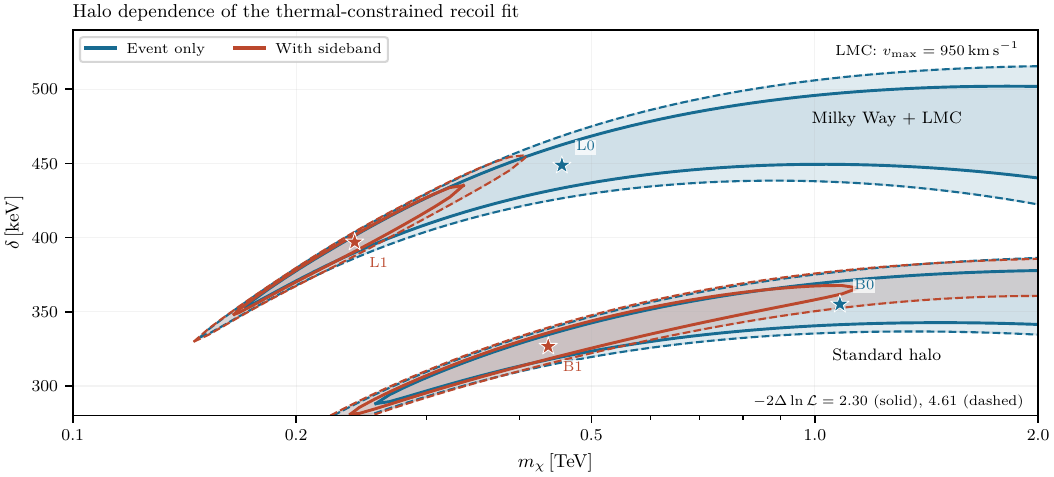}
 \caption{Thermal-constrained mass--splitting fits for the SHM and
 the truncated present-day Milky Way--LMC distribution
 \cite{SmithOrlikEtAl2023}. Colors distinguish the sideband choices,
 solid and dashed contours give conditional likelihood drops 2.30
 and 4.61 from each hypothesis's own maximum. Stars mark B0/B1 and
 L0/L1. The common mass domain begins at 100 GeV. These are
 recoil-only halo comparisons with fixed nuclear inputs. \AIFigureNote}
 \label{fig:lmc-profile}
\end{figure*}

\begin{table*}[t]
\caption{LMC recoil-only thermal benchmarks with the 950-km/s cut.
L0 omits the sideband, L1 includes its zero-count proxy. The
response and count definitions match the B points in
Table~\ref{tab:direct-benchmarks}.}
\label{tab:lmc-benchmarks}
\begin{ruledtabular}
\begin{tabular}{lccccccc}
Point & $\mchi$ [TeV] & $\dm$ [keV] & $g_M$ &
$\mug$ [$10^{-4}\,\GeV^{-1}$] & $N_{\rm ROI}$ &
$N_{\rm sideband}$ & $\tau_2$ [$\mu$s]\\
\midrule
L0 & \LZeroMass & \LZeroDelta & \LZeroG & \LZeroMu & \LZeroROI & \LZeroSB & \LZeroTau\\
L1 & \LOneMass & \LOneDelta & \LOneG & \LOneMu & \LOneROI & \LOneSB & \LOneTau\\
\end{tabular}
\end{ruledtabular}
\end{table*}

\begin{figure*}[t]
 \includegraphics[width=\textwidth]{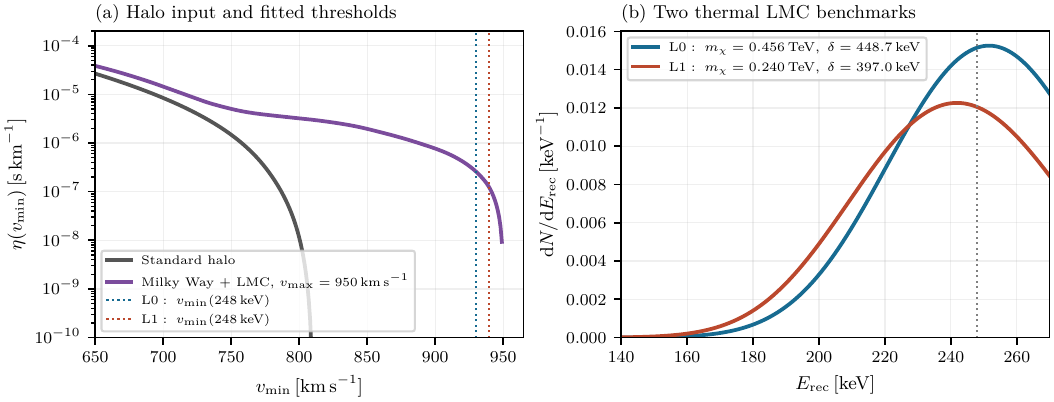}
 \caption{LMC benchmark spectra. Left: normalized inverse-speed
 integrals and the $^{131}$Xe thresholds for L0/L1 at 248 keV.
 Right: selected reconstructed spectra at the same exposure and
 response as Fig.~\ref{fig:decay-spectra}. Their normalization
 follows from the thermal moment. \AIFigureNote}
 \label{fig:lmc-benchmarks}
\end{figure*}

The thermal relation is unchanged. Refitting mass, splitting, and
scale gives the interior maxima in Table~\ref{tab:lmc-benchmarks}.
The enhanced tail permits larger splittings, as shown in
Fig.~\ref{fig:lmc-profile}. L0 predicts $\LZeroSB$ sideband events,
including the empty count moves the maximum to L1 with
$\LOneSB$ events. The 4.61 contour also admits a disconnected
high-mass region, illustrating the weak mass determination from
one event. Figure~\ref{fig:lmc-benchmarks} shows the halo integrals
and resulting spectra.

Lowering $v_c$ to 925 and 900 km/s moves the sideband maxima to
$(\LOneMassCutNineTwentyFive\,\TeV,\LOneDeltaCutNineTwentyFive\,\keV)$
and $(\LOneMassCutNineHundred\,\TeV,\LOneDeltaCutNineHundred\,\keV)$.
These cuts measure sensitivity to the visible endpoint. A statistical
halo uncertainty requires an ensemble of phase-space distributions.

L1 lies below H.E.S.S.'s 300-GeV threshold, with a Fermi NFW
$\gamma\gamma$ ratio $\LOneFermiGG$
\cite{HESS2026Lines,FosterLines2023}. The fits here retain the
recoil-only response. At L1 the lifetime is $\LOneTau\,\mu$s
and characteristic flight is $\LOneLength$ m, so the decay
selection remains relevant. An isotropic speed-only calculation
could extend the G response to this halo, and a physical directional
fit needs its phase-space distribution and the detector response.

%% file: sections/appendix_solar.tex
\section{Solar capture and annihilation}
\label{app:solar}

For Sun-frame speed $u$ at infinity,
$w^2=u^2+v_{\rm esc,\odot}^2(r)$. We obtain the escape speed from
the BS05(AGS,OP) enclosed-mass profile
\cite{BahcallSerenelliBasu2005}, giving approximately 1384 km/s
at the centre and 618 km/s at the surface.
Our first-capture calculation uses stationary nuclei.

The recoil endpoints and capture threshold on isotope $i$ are
\cite{NussinovWangYavin2009,MenonEtAl2010}
\begin{align}
 E_i^\pm&=\frac{\mu_{\chi i}^2w^2}{2m_i}
 \left(1\pm\sqrt{1-\frac{2\dm}{\mu_{\chi i}w^2}}\right)^2,\nonumber\\
 E_C&=\tfrac12\mchi u^2-\dm .
 \label{eq:solar-recoil-endpoints}
\end{align}
Capture requires $E_R>E_C$ and an open inelastic interval.
The subsequent photon removes the excitation, with a small daughter
kick of order $\dm/\mchi$. Writing $E_L=\max(E_i^-,E_C)$, the
optically thin rate is
\cite{Gould1987,CatenaHellstrom2018,BlennowEtAl2015}
\begin{align}
 C_\odot=\frac{\rho_\chi}{\mchi}\sum_i
 &\int_0^{R_\odot}4\pi r^2n_i(r)\,\dd r\nonumber\\
 {}\times&\int_0^\infty\frac{f_\odot(u)}u w^2\,\dd u
 \int_{E_L}^{E_i^+}\frac{\dd\sigma_i}{\dd E_R}\,\dd E_R .
 \label{eq:solar-capture-integral}
\end{align}
An empty recoil interval contributes zero. The cross section
contains the same coherent $O_{1,4,5,6}$ combination as the xenon
calculation.

We include 16 targets:
$^1$H, $^3$He, $^4$He, $^{12}$C, $^{14}$N, $^{16}$O,
$^{20}$Ne, $^{23}$Na, $^{24}$Mg, $^{27}$Al, $^{28}$Si,
$^{32}$S, $^{40}$Ar, $^{40}$Ca, $^{56}$Fe, and $^{58}$Ni
\cite{CatenaHellstrom2018}. BS05 supplies radial H, He, C, N,
and O profiles. Other elements use AGS05 surface abundances
\cite{AsplundGrevesseSauval2005} with the oxygen radial diffusion
factor. The Sun-frame SHM has $v_0=238$ km/s, escape speed
544 km/s, and solar speed 250.6 km/s
\cite{McCabe2014Velocity}. Increasing all quadrature orders
by 50\% changes the four benchmark rates by at most
$2.1\times10^{-4}$. Their scattering probabilities are below
$6\times10^{-6}$ of the focused geometric flux.
Thermal target motion, minor isotopes, and solar-composition
variations remain physical uncertainties.

\begin{table*}[t]
\caption{First capture and conditional annihilation at the fitted
B0/B1/L0/L1 particle parameters. Every row uses the Sun-frame SHM.
The L rows vary particle parameters only. Thermal-core and
uniform-solar-volume annihilation estimates assume the indicated
spatial distributions.}
\label{tab:solar-benchmarks}
\begin{ruledtabular}
\begin{tabular}{lcccccc}
 & $C_\odot$ [$10^{21}$ s$^{-1}$] & Fe fraction & Ni fraction &
$t_\odot\sqrt{C_\odot C_A^{\rm th}}$ &
$\Gamma_A^{\rm th}$ [$10^{21}$ s$^{-1}$] &
$\Gamma_A^{R_\odot}$ [$10^{17}$ s$^{-1}$]\\
\midrule
\input{sections/solar_table_rows.tex}
\end{tabular}
\end{ruledtabular}
\end{table*}

Fe and Ni supply about 99\% of the capture rate
(Table~\ref{tab:solar-benchmarks}). Their coherent charge response
survives on spin-zero isotopes. A spin-dominated xenon rate
therefore leaves an appreciable solar channel.

Neglecting evaporation, the annihilation rate is
\cite{MenonEtAl2010,BlennowEtAl2015}
\begin{equation}
 \Gamma_A=\frac{C_\odot}{2}\tanh^2(t_\odot\sqrt{C_\odot C_A}),
 \qquad C_A=\frac{\sv_{11}}{V_{\rm eff}}.
 \label{eq:solar-annihilation-solution}
\end{equation}
Here $V_{\rm eff}=(\int n\,\dd^3r)^2/\int n^2\,\dd^3r$
and $t_\odot=4.57$ Gyr \cite{BahcallSerenelliBasu2005}.
We use the finite-mass $\gamma\gamma,\gamma Z,ZZ$ rate for
$\sv_{11}$. A thermal Gaussian has
$V_{\rm eff}^{\rm th}=(2\pi)^{3/2}r_\chi^3$ and
$r_\chi^2=3k_BT_c/(2\pi G\rho_c\mchi)$.
It brings B0/B1 close to equilibrium. 

\begin{figure*}[t]
 \includegraphics[width=.97\textwidth]{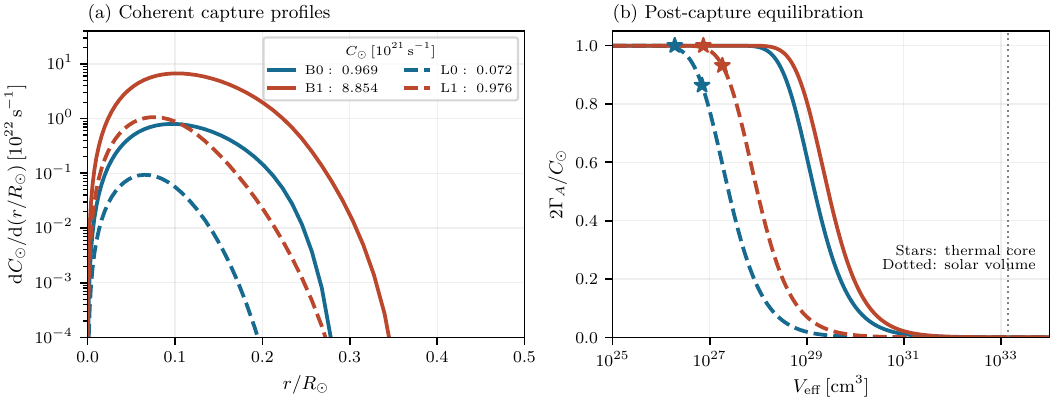}
 \caption{Solar capture at the four fitted particle points with a
 common Sun-frame SHM. Left: radial first-capture density.
 Right: annihilation fraction $2\Gamma_A/C_\odot$ versus effective
 volume, stars mark the thermal-core assumption. Solid curves are
 B0/B1 and dashed curves L0/L1. Solar inputs follow
 Refs.~\cite{BahcallSerenelliBasu2005,AsplundGrevesseSauval2005}. \AIFigureNote}
 \label{fig:solar-capture}
\end{figure*}

Figure~\ref{fig:solar-capture} displays this volume dependence.
Inelastic rescattering shuts off as orbits fall below threshold
\cite{NussinovWangYavin2009,MenonEtAl2010,BlennowEtAl2015}.
The actual captured distribution must follow from orbital evolution.
UV-induced elastic cooling can change that distribution while
leaving the LZ transition dominant.

The propagated neutrino flux is
\begin{equation}
 \frac{\dd\Phi_{\nu_\alpha}}{\dd E}
 =\frac{\Gamma_A}{4\pi D_\odot^2}
 \sum_f B_f\left.\frac{\dd N^f_{\nu_\alpha}}{\dd E}
 \right|_{\rm propagated}.
 \label{eq:solar-neutrino-flux}
\end{equation}
Propagation includes oscillation, absorption, and regeneration
\cite{BlennowEdsjoOhlsson2008}. The minimal model supplies neutrinos
through $Z$ decay, with mean $Z$ multiplicity
$B_{\gamma Z}+2B_{ZZ}\to2s_W^2$. An IceCube comparison
\cite{IceCubeSolar2025} must use these yields and the orbital
distribution. Additional annihilation channels change both factors.

%% file: sections/solar_table_rows.tex
B0 & 0.969 & 0.928 & 0.067 & 24.32 & 0.485 & 0.3870 \\
B1 & 8.854 & 0.923 & 0.066 & 17.04 & 4.427 & 6.7454 \\
L0 & 0.072 & 0.915 & 0.085 & 1.65 & 0.031 & 0.0005 \\
L1 & 0.976 & 0.924 & 0.075 & 2.02 & 0.455 & 0.0256 

%% file: sections/appendix_decay_selection.tex
\section{Transport and line-limit details}
\label{app:decay-selection}

\subsection{Geometry and speed averages}

Write the isotope rate as
$K_{\eta,T}\eta+K_{\perp,T}\eta_\perp$, with dimensionless
velocities. An isotropically averaged acceptance $A(v_2)$ enters as
\begin{equation}
 \int_{v_{\min,T}}^{v_{\max}}\frac{f_1(v)}v\,\dd v\,
 [K_{\eta,T}+K_{\perp,T}(v^2-v_{\min,T}^2)]A(v_2).
 \label{eq:weighted-speed-kernel}
\end{equation}
This retains the response-weighted outgoing-speed distribution.
The angular average is a separate approximation: the fast halo
population has a preferred incident direction.

For vertex $(x,y,z)$, direction $\bm n$, $a=n_x^2+n_y^2$,
and $b=xn_x+yn_y$, the cylindrical exit distance is
\begin{align}
 D_r&=\frac{-b+\sqrt{b^2+a(R^2-r^2)}}a,\nonumber\\
 D_z&=\begin{cases}(H-z)/n_z,&n_z>0,\\-z/n_z,&n_z<0,\end{cases}
 \qquad D=\min(D_r,D_z).
 \label{eq:cylinder-distances}
\end{align}
A vanishing directional component gives an infinite distance to
that surface. The volume and event-position averages integrate
$e^{-D/\lambda_2}$ separately. The survival curve in
Fig.~\ref{fig:decay-geometry} is
\begin{equation}
 S(L|E_{\rm obs})=\langle e^{-L/\lambda_2}\rangle_{E_{\rm obs}},
 \qquad \langle\ell_2\rangle=\int_0^\infty S(L|E_{\rm obs})\,\dd L .
 \label{eq:mixture-survival}
\end{equation}
The average includes recoil energy, isotope, incident speed, and
the candidate's resolution and scale pull.

Doubling velocity nodes changes benchmark observables by at most
$\GeometryVelocityError$ relatively. Independent geometry samples
change the fixed-length probability by at most
$\GeometryProbabilityError$ absolutely. Spectral integrals recover
the counts to $2\times10^{-6}$ relatively.

The true fiducial stand-off ranges from 8.0 to 18.2 cm
\cite{LZ2026HighER}. Uniform-cylinder variations over those values
give B0 volume escape probabilities 0.386--0.355, compared with
0.377 at the nominal 10.7 cm. B1 gives 0.389--0.358, compared with
0.379. The exposure remains fixed.
A directional calculation uses
$\hat{\bm v}\cdot\hat{\bm q}_T=v_{\min}/v$ and
$\bm v_2\simeq\bm v-\bm q_T/\mchi$, with $\bm q_T$ toward the
nucleus. Its orientation relative to the chamber requires the
event azimuth and live-time-dependent halo wind.

\subsection{Photon escape, return, and vetoes}

For a decay at $\bm x_2=\bm x+\ell_2\bm n_2$, let
$L_\gamma^{\rm TPC}$ be the forward photon-ray length within the
active cylinder, including re-entry after an external decay.
A first-interaction transport estimate gives
\begin{equation}
 P_{\gamma,0}=\left\langle e^{-L_\gamma^{\rm TPC}/\ell_\gamma}\right\rangle,
 \qquad \ell_\gamma^{-1}=\rho_{\rm LXe}(\mu/\rho)_{\rm Xe}.
 \label{eq:photon-transparency}
\end{equation}
We take photons isotropic in the excited-state rest frame and
use total xenon attenuation coefficients, 0.1797 and
$0.1223\,\mathrm{cm^2\,g^{-1}}$ at 300 and 400 keV
\cite{NISTXenonAttenuation}. External material is transparent in
this geometry test, every TPC interaction is treated as visible.

At B0/B1, the probability of no photon interaction in the TPC is
$\BZeroPhotonTransparent$ and $\BOnePhotonTransparent$, compared with
$\BZeroEventEscape$ and $\BOneEventEscape$ for parent exit alone.
Photons escaping after internal decays add
$\BZeroPhotonEscapeAdded$ and $\BOnePhotonEscapeAdded$ in absolute
acceptance, while returning photons subtract
$\BZeroPhotonReturnRemoved$ and $\BOnePhotonReturnRemoved$.
This calculation shows a material correction to a TPC-exit model.
It has yet to be included in a full response refit.

An unbounded decay clock gives
$P(t_2<t)=1-e^{-t/\tau_2}$, but a measured delay distribution
also carries spatial, energy, and pulse-finding weights.
The prompt and delayed Skin/outer-detector windows
\cite{LZ2026HighER}, Compton energy loss, reverse-field
scintillation, and merged pulses all influence migration.
Their combined acceptance requires a signal-injection study.

\subsection{Line recasting}

The Fermi observed curve is extracted from
Ref.~\cite[Fig.~4]{FosterLines2023}, retaining 455 vector
vertices and logarithmic interpolation. At 437.37 GeV it gives
a native-density limit $3.36\times10^{-27}\,\cms$.
The H.E.S.S. extraction retains 61 vertices from about 306 GeV
to 63.4 TeV, its minimum,
$9.787\times10^{-29}\,\cms$ at 437.0 GeV, agrees with the
published $9.8\times10^{-29}$ anchor near 440 GeV
\cite{HESS2026Lines}. Benchmark comparisons interpolate within
these domains.

The Fermi boundary in Fig.~\ref{fig:combined-constraints} uses
$\gamma\gamma$ alone. For an individual $\gamma Z$ feature,
matching its flux to a two-photon limit at $E_{\gamma Z}$ gives
\begin{equation}
 \sigma_{\gamma\gamma}^{\rm equiv}(E_{\gamma Z})
 =\tfrac12\sv_{\gamma Z}(E_{\gamma Z}/\mchi)^2 .
 \label{eq:individual-gammaZ-recast}
\end{equation}
For unresolved features we use the photon-weighted sum in
Eq.~\eqref{eq:line-rate}, including the H.E.S.S. comparison.
A joint energy-response likelihood would refine the doublet
treatment. Density rescalings retain each analysis's own spatial
profile, and all translated moment limits assume the point
hypercharge hard amplitude.

%% file: sections/appendix_secluded.tex
\section{A free moment over 0.1--100 TeV}
\label{app:secluded}

\begin{figure*}[t]
 \includegraphics[width=\textwidth]{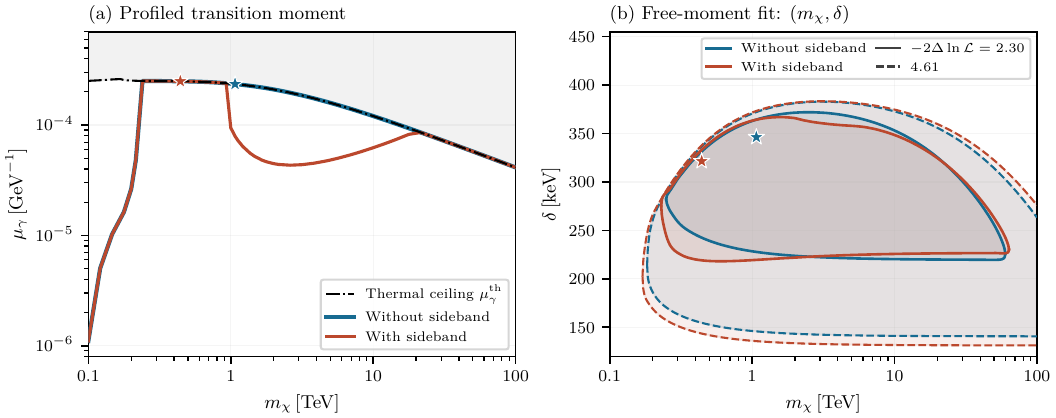}
 \caption{Free-moment extension of Fig.~\ref{fig:thermal-lz-target}.
 Left: the best-fit moment at each mass, with splitting and energy
 scale profiled, compared with the thermal ceiling. Right:
 mass--splitting support after profiling the moment below that ceiling
 and the scale pull. Blue omits the sideband, orange includes its
 empty proxy. Both use TPC-exit selection. Solid and dashed contours
 denote conditional likelihood drops 2.30 and 4.61, with uncalibrated
 coverage. Stars mark the global maxima. The high-mass thermal
 ceiling assumes extrapolation of the point hypercharge amplitude. \AIFigureNote}
 \label{fig:wide-free-dipole}
\end{figure*}

Additional annihilation can set the abundance while the dipole
controls recoil and decay~\cite{PospelovRitzVoloshin2008,BatellPospelovRitz2009}.
We retain the full local dark-matter density and profile the
escape-selected likelihood over
\begin{align}
 &0<\mug\leq\mug^{\rm th}(\mchi),\nonumber\\
 &0.1\leq\mchi/\TeV\leq100,\qquad 1\leq\dm/\keV\leq700.
 \label{eq:wide-free-domain}
\end{align}
The thermal ceiling assumes a symmetric, common-temperature
relic with additional positive annihilation rates in a standard
radiation bath. Its calculation uses $\dm=0$, retaining finite-mass
kernels and the temperature-dependent equation of state.
Keeping $\dm=700$ keV changes the abundance by less than
$\WideSplittingError$ at the checked points.
At high mass the $\mug^4\mchi^2$ term lowers the ceiling
approximately as $\mchi^{-1/2}$.
At 100 TeV, $2\mchi\mu_Y=\WideHardExpansionHundred$:
This part of the scan is a conditional hard-operator extrapolation.

Figure~\ref{fig:wide-free-dipole} shows the enlarged parameter space.
Both global maxima still reach the ceiling, within $0.04\%$ in mass
of G0/G1 after setting the relic splitting to zero.
At 6.8 TeV, the sideband profile selects
$\dm=\WideProfileSixDelta$ keV and
$\mug=\WideProfileSixMu\,\GeV^{-1}$, with
$\WideProfileSixROI$ selected and $\WideProfileSixSideband$ sideband
events. Its likelihood drop is $\WideWOneDropSix$, mean flight
$\WideProfileSixFlight$ m, and point-dipole line rate
$\WideProfileSixLine\,\cms$, or $\WideProfileSixHess$ of the
H.E.S.S. Einasto limit. The drops at 14 and 100 TeV are
$\WideWOneDropFourteen$ and $\WideWOneDropHundred$.
One event therefore leaves a broad heavy-mass alternative.
The abundance-setting interaction and any independent Rayleigh
amplitude determine its remaining indirect constraints.

For comparison, Asadi, Batz, Fox, Homiller, and Kribs
\cite{AsadiEtAl2026Magnetic} fix a dimensionless dark-baryon moment,
$|g_F|=2\sqrt3$, corresponding to $\mug=e|g_F|/(2\mchi)$
and our $g_M=4\sqrt3$. They quote preferred points near
$(6.8\,\TeV,300\,\keV)$ and $(14\,\TeV,280\,\keV)$ for their
SHM and LMC choices. Their six-bin response and $\mug\propto1/\mchi$
The trajectory differs from the free physical moment profiled here.
These choices lead to different mass preferences within a shared
recoil operator. Their confining abundance and hard annihilation
amplitudes also require a separate calculation.
The companion notes retain the matched-input comparison.